\documentclass[lettersize,journal]{IEEEtran}
\usepackage{amsmath,amsfonts}
\usepackage{algorithmic}
\usepackage{array}
\usepackage{textcomp}
\usepackage{stfloats}
\usepackage{url}
\usepackage{verbatim}
\usepackage{graphicx}
\def\BibTeX{{\rm B\kern-.05em{\sc i\kern-.025em b}\kern-.08em
    T\kern-.1667em\lower.7ex\hbox{E}\kern-.125emX}}
\usepackage{balance}
\usepackage{subfigure}
\usepackage{makecell}
\begin{document}
\title{Point Cloud Objective Quality:\\ Benchmarking Features and Quality Evaluation
}
\author{Joao~Prazeres,~\IEEEmembership{Student Member,~IEEE}, Rafael~Rodrigues,~\IEEEmembership{Student Member,~IEEE},
Manuela~Pereira,~Antonio~M.~G.~Pinheiro,~\IEEEmembership{Senior Member,~IEEE}
\thanks{Joao Prazeres, Rafael Rodrigues and Antonio M. G. Pinheiro are with Instituto de Telecomunicacoes \& Universidade da Beira Interior, Portugal.}
\thanks{Manuela Pereira is with NOVA LINCS \& Universidade da Beira Interior, Portugal.}
\thanks{This work is funded by FCT/MECI through national funds and when applicable co-funded EU funds under UID/50008: Instituto de Telecomunicações}}

\markboth{IEEE TRANSACTIONS ON CIRCUITS AND SYSTEMS FOR VIDEO TECHNOLOGY}%
{How to Use the IEEEtran \LaTeX \ Templates}

\maketitle

\begin{abstract}
Full-reference point cloud objective metrics are currently providing very accurate representations of perceptual quality. These metrics are usually composed of a set of features that are somehow combined, resulting in a final quality value.
In this study, the different features of the best-performing metrics are analyzed.
For that, different objective quality metrics are compared between them, and the differences in their quality representation are studied.
This provided a selection of the set of metrics used in this study, namely the
point-to-plane, point-to-attribute, Point Cloud Structural Similarity, Point Cloud Quality Metric and Multiscale Graph Similarity. The features defined in those metrics are examined based on their contribution to the objective estimation using recursive feature elimination. To employ the recursive feature selection algorithm, both the support vector regression and the ridge regression algorithms were employed.
For this study, the Broad Quality Assessment of Static Point Clouds in Compression Scenario database was used for both training and validation of the models.
According to the recursive feature elimination, several features were selected and then combined using the regression method used to select those features.
The best combination models were then evaluated across five different publicly available subjective quality assessment datasets, targeting different point cloud characteristics and distortions.
It was concluded that a combination of features selected from the Point Cloud Quality Metric, Multiscale Graph Similarity and  PSNR MSE D2, combined with Ridge Regression, results in the best performance. This model leads to the definition of the Feature Selection Model.
\end{abstract}


\section{Introduction}
\IEEEPARstart{T}he current technological evolution is experiencing an increasing need for 3D data formats~\cite{Schwarz2019}. Point clouds became one of the most popular methods for representing volumetric data and consist of a set of Cartesian coordinates $(x, y, z)$. Each coordinate may have an associated list of attributes, such as an RGB component, reflective information, physical sensor information, or normal vectors. 
Typically, point clouds contain large amounts of information for the representation of objects or scenes.
Hence, most applications using point cloud representation models will benefit from powerful coding solutions that provide means for efficient data handling, notably lossy encoding solutions. 
Such solutions might induce significant visual distortions~\cite{GPA2023}, that need to be accurately measured by reliable models that evaluate the quality of the decoded point cloud content. 
The most reliable quality measures result from suitable subjective quality tests that require careful planning and are very time-consuming~\cite{SunWeiJSTSP2023}. Therefore, it is crucial for developers to have reliable objective quality measures for point clouds, as they are necessary for evaluating and improving new coding solutions.

Point cloud quality models exhibit significant differences compared to those observed in conventional 2D images, where pixels are situated in a rigid grid with no empty spaces between them. The points are also unevenly distributed in a volumetric space, making it challenging to develop efficient point cloud quality evaluation models.
Recently, many objective quality metrics, ranging from full-reference, reduced-reference and no-reference, were developed for point clouds. The full-reference metrics, which directly compare the reference and distorted content, usually provide the best performance~\cite{PRAZERESSPIC2024}. 

Machine learning has been increasingly used in quality models.
Successful learning-based metrics tend to be based on 
learning models, where a set of features is classified as a given quality with a common classifier.
For instance, VMAF~\cite{VMAF}, uses a trained Support Vector Regression (SVR)~\cite{SVR2015} model using several objective quality metrics for image and video quality estimation.

Most of the recent point cloud quality assessment metrics rely on a set of features developed to deal with both the geometry and color of the point clouds information. These features are further combined, resulting in a final quality measure.  

This paper  aims to report the results of a study developed in  the context  of the JPEG PLENO Learning-Based Point Cloud Coding activity~\cite{FSMStudyJPEG}.
Hence, the main contributions of this paper are as follows:
\begin{itemize}
    \item An analysis of the contribution of the features defined in different metrics for the final quality estimation.
    \item Analysis of different combinations of features, leading to the definition of a model that combines those features using the best regression method, named Feature Selection Model (FSM).
\end{itemize} 

This work is motivated by the VMAF~\cite{VMAF} approach for image and video quality assessment, where multiple quality features are combined to obtain a quality estimation.
It should be noted that the main goal of this work is not the development of new quality features.
The contribution of different features defined in different metrics for the perceptual quality estimation is evaluated. Then, the best combination is defined in order to provide a more reliable quality measure.
The fact that the different features of the metrics provide different quality representations is the main motivation for this study. Analyzing the behavior of these features and how relevant their contribution is to quality prediction will be useful for researchers in this domain and in the development of new objective methods.

The initial metrics considered for this study were the point-to-point (PSNR MSE D1)~\cite{Dtian}, point-to-plane (PSNR MSE D2)~\cite{Dtian}, point-to-attribute (PSNR MSE YUV)~\cite{PSNRYUV}, Point Cloud Structural Similarity (PointSSIM)~\cite{AlexiouPointSSIM}, Point Cloud Quality Metric (PCQM)~\cite{MeynetPCQM}, and Graph Similarity (GraphSIM)~\cite{QiYangGraphSIM2022} along with its expanded Multiscale Graph Similarity (MS-GraphSIM)~\cite{MSGraphSIM}.
In this work image metrics were not considered, although they can be applied for point cloud objective quality evaluation~\cite{PRAZERESSPIC2024,ak2023basics,alexiou2019a,PSNRYUV,LazzarottoMMSP2021,Javaheri_projection2022,TOPIQ2024}. However, these metrics depend on the visualization directions of the point clouds, leading to some instability. Furthermore, some recent works~\cite{ak2023basics,PRAZERESSPIC2024}  seem to indicate that the most recent point cloud metrics tend to provide better performance.

\begin{figure}[!t]
    \centering
    \includegraphics[width=\linewidth]{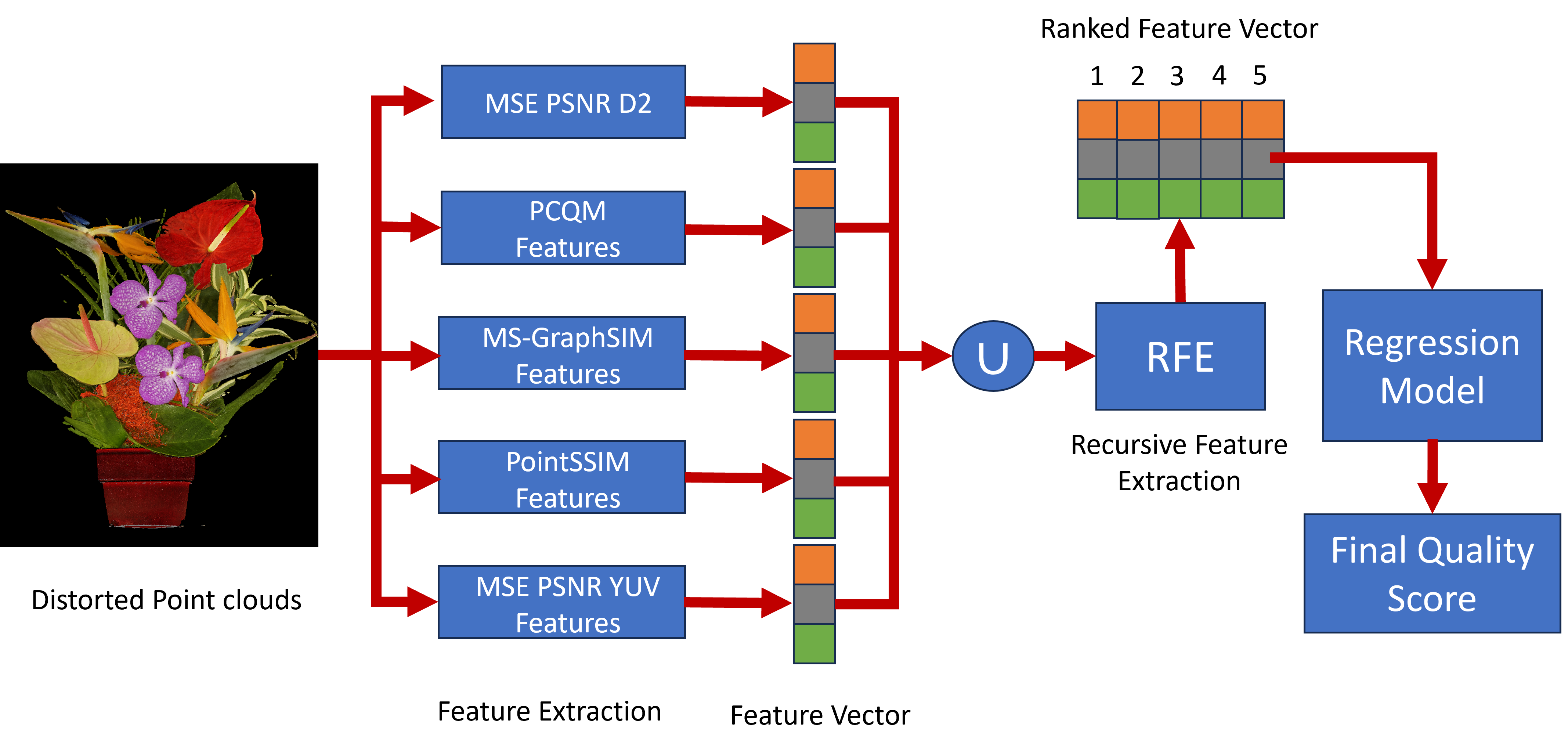}
    \caption{feature extraction and regression framework. Extracting the features results in a vector containing all computed features. RFE analyzes the feature importance to create a Ranked Feature Vector. Finally, the regression model computes the final quality scores.}
    \label{fig:FSMFramework}
\end{figure}

The assessment of the quality features  uses Recursive Feature Elimination (RFE)~\cite{RFE}.
Furthermore, the Support Vector Regression (SVR)~\cite{SVR2015} and Ridge Regression (RR)~\cite{RidgeRegression} algorithm were used for the final quality estimation using the selected features. 
The Random Forest Regression~\cite{cutler2012random} and AdaBoost~\cite{AdaBoost1997} were also considered, but the results are not reported as they resulted in worst performance.

The Broad Quality Assessment of Static Point Clouds in Compression Scenario (BASICS) training dataset~\cite{ak2023basics} is used for the analysis of the contribution of each metrics features.
The main target of this study is to assess the quality of colored static point cloud coding solutions. The BASICS dataset was selected for training and validation because it is the largest annotated publicly available point cloud dataset containing distortions created by conventional codecs and learning based-codecs.

Figure \ref{fig:FSMFramework} shows the framework of this study. Features from the aforementioned metrics will be extracted from both the reference and distorted point clouds, resulting in a feature vector. That vector will be analyzed using RFE with a regression method to obtain a ranking of the most important features. Finally, the most important features are selected and used as input to the regression method used to rank the features, leading to a quality estimation.

The remainder of this paper is organized as follows: Section \ref{sec:relatedWork} reports on the state-of-the-art of point cloud quality evaluation under coding distortions, as well as the several existing objective quality metrics. It also includes a brief description of the most relevant point cloud codecs. The considered metrics for this evaluation are described in Section \ref{sec:Metrics}. The regression method, training dataset, and the method used for feature selection are described in Section \ref{sec:FeatureAnalysis}. Section \ref{sec:Validation} describes the benchmarking of several combinations of features and provides details of the considered datasets.  
Finally, Section \ref{sec:conclusions} summarizes the main conclusions.

\section{Related Work}\label{sec:relatedWork}
In this section, a summary of several works on subjective and objective quality assessment is presented, as well as some of the most relevant models for point cloud coding.

\subsection{Subjective quality evaluation}
Given the need to evaluate point cloud compression solutions, extensive research has been conducted on the subjective quality assessment of point clouds that provide quality models for the evaluation of different coding methodologies and their respective parametrization.

Several studies were carried out to establish quality models for geometry-only~\cite{Alexiou2018a,alexious2018point}, graph-based~\cite{Javaheri2017b}, and projection-based~\cite{Silva2019a} codecs.
Honglei \textit{et al.} conducted subjective evaluations to study coding solutions that would become the early stages of MPEG standards, namely the codecs V-PCC and G-PCC~\cite{Su2019a}.
A study on those coding solutions before their final standardization was also reported~\cite{alexiou2019a}.

Objective and subjective quality evaluations were conducted to assess the performance of MPEG standards using a 2D setup~\cite{icip2020}. It was concluded that V-PCC was the best-performing codec. Those solutions were compared with Draco(https://github.com/google/draco) and RS-DLPCC~\cite{GuardaRSDLPCC} using a 2D display~\cite{EI2022}, followed by a study where the 2D display was replaced by a 3D stereoscopic visualization setup~\cite{ICIP2022}.

Subjective evaluation studies using virtual or augmented reality (VR/AR) environments have also been reported~\cite{AlexiouMMSP2017, ViolaVr, MekuriaVR,Alexiou:277378}.
Moreover, a subjective quality assessment study targeting learning-based coding solutions~\cite{PrazeresACM2022} was conducted, using a set of six point clouds depicting both objects and landscapes encoded with three deep-learning-based codecs. The tested learning-based solutions showed competitive results when compared with G-PCC.

JPEG Pleno Point Cloud Coding activity also reported a subjective study using state-of-the-art codecs prior to the launch of its call for proposals~\cite{PerryEUVIP2022}. The aim was to evaluate state-of-the-art point cloud solutions, analyze the stability of subjective quality assessment models, and evaluate the performance of objective metrics.
The call for proposals results were presented by Prazeres \textit{et al.}~\cite{PrazeresICASSP2023}. Three deep-learning based solutions capable of encoding both geometry and color were evaluated, and the best performing solution became the base for the JPEG Pleno Point Cloud Learning-based Verification Model~\cite{VM-CD}
Another paper~\cite{LIUQI2022} reports the subjective evaluation of several types of distortion in a large point cloud database, with an extensive metrics benchmarking study.
A large benchmarking study was also conducted by Prazeres \textit{et al.}~\cite{PRAZERESSPIC2024}, where several full-reference, reduced-reference, and no-reference metrics were evaluated. It revealed that the no-reference metrics do not achieve an acceptable representation of the subjective results.
Moreover, Perry \textit{et al.}~\cite{MMSP2021} reported on a subjective study using crowdsourcing methodologies. Participants were able to either download the subjective evaluation content or access an online server and perform the evaluation in a web browser. The two types of subjective evaluation revealed a very high level of statistical similarity.

\subsection{Objective quality evaluation}\label{sec:ObjMetrics}

Objective quality evaluation metrics are critical in the development of coding methods, as they do not require long and costly subjective quality evaluation. These metrics are usually developed to provide the best possible representation of the subjective evaluation. Some metrics only represent a measure of the signal fidelity, but in those cases, the representation of the subjective results is frequently limited.
This section analyzes the state-of-the-art in the objective quality assessment of point clouds.

Point cloud quality evaluation metrics can be divided according to the type of information considered: 
\begin{enumerate}
    \item Geometry only, that only considers the geometry of the point cloud,
    \item Color only that only consider the color attributes of the point cloud, and
    \item Geometry and color, that considers both the geometry and color attributes of the point cloud.
\end{enumerate}

Full-reference metrics (where the distorted data is compared with the original data) have been the main focus of the objective quality assessment of point clouds. Early point cloud metrics, namely point-to-point (PSNR MSE D1), point-to-plane (PSNR MSE D2)~\cite{Dtian}, and point-to-mesh~\cite{p2mesh}, evaluate point cloud quality using only geometry information. PSNR MSE D1 measures the geometric distance between corresponding points in the reference and the distorted point cloud, while the PSNR MSE D2 metric projects the distance vector onto the surface normal orientation in order to evaluate point cloud quality. Point-to-mesh measures the distance between points in the distorted point cloud and a mesh reconstruction of the reference point cloud. Both the PSNR MSE D1 and PSNR MSE D2 metrics have been widely used in point cloud benchmarking~\cite{icip2020,ICIP2022,AlexiousQoMEX2017,PrazeresEUVIP2022,JAICMEW2017}, but were shown to be quite unstable in predicting point cloud quality. To tackle this issue, several studies have been conducted to find the best features for point cloud quality assessment. The plane-to-plane metric~\cite{AlexiouPl2Plane} adopts the angular differences between the point cloud normal vectors for quality estimation. Javaheri \textit{et al.}~\cite{JavaheriHauss2020} proposed a method based on the Haussdorff distance to predict geometry distortions. The same authors also developed a full-reference metric that uses the \textit{Mahanobilis} distance to assess color and geometry distortions~\cite{javaheri2021pointtodistribution}. Javaheri \textit{et al.}~\cite{Javaheri_projection2022} also proposed a joint geometry and color projection-based full reference metric. 

For quality assessment of point clouds, attributes are of the utmost importance. To include color information in objective quality models, Meynet \textit{et al.} proposed the Point Cloud Quality Metric (PCQM)~\cite{MeynetPCQM}. It uses a weighted linear combination of curvature and color information to predict the visual quality of a distorted point cloud.
The Graph Similarity (GraphSIM)~\cite{QiYangGraphSIM2022} metric extracts geometric keypoints from the point cloud and then uses graph similarity to evaluate the distortions in the point clouds. This metric was later extended by Zhang \textit{et al.}~\cite{MSGraphSIM} with the Multiscale Graph Similarity (MS-GraphSIM), where the GraphSIM features are computed on different scales.
The Point Cloud Structural Similarity (PointSSIM) metric~\cite{AlexiouPointSSIM} is inspired by the Structural Similarity Index Measure (SSIM)~\cite{SSIM} developed for static images and computes the similarity between geometry, normal vector, color, and curvature features. A 3D to 2D projection metric was proposed by Qi Yang~\cite{projectionYANG}. The Transformational Complexity Based Distortion Metric~\cite{zhang2023tcdm} quantifies point cloud quality as the complexity of transforming it into its corresponding reference.
Color-only methods have also been researched in the literature. The color Histogram~\cite{IreneHistogram} metric extracts color features from the reference and distorted point clouds and compares the resulting color histogram for each point cloud.
The point-to-attribute metric (PSNR MSE YUV)~\cite{PSNRYUV} is based on the error of the color values between the identified point in the reference and the distorted point cloud. The identification process is conducted using the nearest neighbor algorithm, and an individual error for each color channel is computed for the identified points based on the Euclidean distance. The overall PSNR is computed by weighting each color channel as $Y:U:V = 6:1:1$ respectively. A recent paper~\cite{watanabe2025full} extracts five different features, based on PSNR MSE D1, PSNR MSE D2, PointSSIM and GraphSIM. The final feature is based on the number of points of the point cloud. The features are used to train an SVR model.


There are also reduced reference metrics developed for point cloud quality estimation, where a set of features extracted from the distorted and reference point clouds are compared.
Viola \textit{et al.}~\cite{ViolaPCMRR} proposed to extract a small set of geometry, color, and normal features that are used to predict the visual quality of the content under evaluation. Then, a linear optimization algorithm is used to find the best weights for each extracted feature. Another reduced reference metric was presented by Zhou \textit{et al.}~\cite{ZhouRRCap}, which considers content-oriented similarity and statistical correlation measures to assess point cloud quality.

There are still few works on no-reference metrics in the literature, and most are learning-based. The ResSCNN~\cite{LiuResSCNN2023} model is based on hierarchical feature extraction using sparse convolution blocks with residual connections. PQA-Net~\cite{PQA-NEtLiu} was also proposed as a learning-based method using multiview-based feature extraction and fusion for both distortion type identification and quality prediction.
Zhan \textit{et al.}~\cite{ZhanSVR2022} trained an SVR with 3D natural scene statistics and entropy features from the geometric and texture domains to evaluate the quality of both point clouds and meshes.
Recently, Marouane \textit{et al.} presented several works on learning-based no-reference metrics, based on a graph edge convolutional network~\cite{TlibaICASSP2023}, on cross-correlation of local features~\cite{MarouaneACM2022}, and on a transformer model using a PointNet backbone~\cite{qi2017pointnet} for feature embedding~\cite{tliba2023quality}.
Yang \textit{et al.} presented a no-reference metric based on 3D to 2D projections and using unsupervised adversarial domain adaptation~\cite{yang2022ITPCQA}. Xiong \textit{et al.}~\cite{XiongICASSP2023} presented a metric based on a point structural information (PSI) network ($\psi$-Net). A PSI module maps the geometric and color structure, and then a dual-stream network introduces distortion-related features. Zicheng \textit{et al.}~\cite{ZhangZichengGMS} proposes a methodology to assess the quality of point clouds using projections. Furthermore, Wei \textit{et al.}~\cite{WeiBQAACMToMM2024} extracts geometry, color and normal based features, then uses a combination of regression based models to evaluate quality.

\subsection{Point cloud coding}
The most traditional point cloud coding models are based on the octree pruning method~\cite{rusu2011a}.
Recently, MPEG defined Geometry-Based Point Cloud Compression (G-PCC)~\cite{VPCCandGPCC} for static point clouds, based on the octree point cloud representation.
G-PCC also specifies the possibility of using the trisoup method, which is based on surface reconstruction for geometry compression.
Furthermore, the point cloud attributes are compressed using either RAHT~\cite{Queiroz2016a} or the lifting transform~\cite{VPCCandGPCC}.

Another approach to point cloud coding relies on encoding the point cloud projections, which can be coded using image or video codecs. MPEG also explored this approach, resulting in the Video-Based Point Cloud Compression (V-PCC)~\cite{VPCCandGPCC} for dynamic point clouds. V-PCC uses High Efficiency Video Coding (HEVC), or more recently, Versatile Video Coding (VVC), to encode 2D projections of a given point cloud. Despite being developed for dynamic point clouds, its intra-coding has revealed to be the most efficient for static point cloud coding~\cite{EI2022,icip2020}.

Following the good performance in image coding, several machine learning-based coding solutions for point clouds have been proposed recently~\cite{8954537,GuardaRSDLPCC,Jianqiang-PCGC,Jianqiang-PCGCv2,quach2019-geocnn,quach2020improved,GuardaDL,sparsePCGC,GuardaTMM2023}.
Learning-based encoders usually cause distortions that are quite different from those caused by conventional codecs. A common distortion is the appearance of empty spaces in the point cloud geometry. These type of codecs are becoming more and more common, mainly due to the popularity of deep-learning technology~\cite{VALENZISE2023}. Because of their rising popularity, it is crucial that objective quality models are able to accurately benchmark these kinds of solutions.

\section{Metrics description}\label{sec:Metrics}
In the following section, a brief description of the selected metrics for this study is presented. 

\subsection{Point-to-Plane - PSNR MSE D2~\cite{Dtian}}\label{D2}
The PSNR MSE D2 metric considers the projection of the error vector $\overrightarrow{v}^{a_i}_{b_k}$ along the surface normal of the nearest neighbor $a_i$ ($N_{a_i}$). 
This vector distance can be computed using the Mean Square Error (MSE) or the Hausdorff distance.
Typically, the Hausdorff distance achieves a worst representation of the subjective quality. 
Hence, in this work, only the MSE is used for the PSNR MSE D2 metric computation.

The MSE considering the projected errors is given by $E(a_i,b_k)=\left|{\overrightarrow{v}^{a_i}_{b_k} \cdot N_{a_i}}\right|$ for each point, followed by its mean value computation. Then, the PSNR is computed using: $PSNR=10 \log_{10}\left(\frac{3*peak^2}{MSE} \right)$,
where \textit{peak} is the geometric resolution of the model $(2^{\mbox{bit depth}}-1)$.
The implementation of this metric provided by MPEG\footnote{http://mpegx.int-evry.fr/software/MPEG/PCC/mpeg-pcc-dmetric/tree/master} was used. Furthermore, the normals were computed with quadratic fitting using CloudCompare\footnote{https://www.danielgm.net/cc/} including points within a radius of 20~\cite{JPEGCTC}.

\subsection{Point-to-attribute - PSNR MSE YUV~\cite{PSNRYUV}}
This metric is based on the error of the color values between the identified point in the reference and the distorted point cloud. The identification process is conducted using the nearest neighbor algorithm. An individual error is computed for the identified points based on the Euclidean distance.
For color attributes, MSE or Hausdorff distance is calculated for the three components, with an RGB to YCbCr conversion being made~\cite{color}.
In this work, the MSE is used, because it usually results in a better quality representation than the Hausdorff distance.
The PSNR value, based on MSE, is computed using: $\mbox{PSNR}=10\log_{10} \frac{p^2}{MSE}$,
with \textit{peak} being 255, considering that all the color attributes of the tested point clouds have a bit depth of 8.
The metric is then computed symmetrically. The final value for each color channel is the maximum between the two computations. The final value for the metric is the PSNR value for each color channel, computed by equation (\ref{eq:PSNRColor}),
\begin{equation}\label{eq:PSNRColor}
PSNR_{Color}=\frac{6PSNR_{Y}+PSNR_{C_{b}}+PSNR_{C_{r}}}{8}
\end{equation}
where $PSNR_{Y}$, $PSNR_{C_{b}}$ and $PSNR_{C_{r}}$ are the $PSNR$ values computed for the $Y$, $C_b$ and $C_r$ respectively~\cite{PSNRYUV}.

 The available MPEG implementation was used\footnote{http://mpegx.int-evry.fr/software/MPEG/PCC/mpeg-pcc-dmetric/tree/master}.

\subsection{Point Cloud Structural Similarity~\cite{AlexiouPointSSIM}}\label{sec:PointSSIM}
This metric extracts features to quantify the statistical dispersion of quantities that characterize the local topology and appearance of the point cloud.
Neighbors around every point of a model are selected to capture local properties.
Quantities to reflect local properties are computed, considering four different attributes, notably geometry information, normal vectors, curvature values, and texture information.
For the feature extraction, dispersion statistics are computed using one of the available estimators, namely the median $(m)$, variance $(\sigma^2)$, mean absolute deviation $(\mu_{AD})$, median absolute deviation $(m_{AD})$, coefficient of variation $(COV)$, and quartile coefficient of dispersion $(QCD)$.
The estimators will be applied over a number of $K$ nearest neighbors.
The perceptual quality prediction is based on the feature similarity values extracted from the reference point cloud ($X$) and the distorted point cloud ($Y$). Each neighborhood of $Y$ is associated with a neighborhood of $X$.
Then the similarity is measured as the relative difference between the corresponding feature values ($F_X$ and $F_Y$), with $\epsilon$ being an arbitrary small value, in order to avoid undefined operations.
\begin{equation}\label{SY}
     S_Y(p)=\frac{|F_X(q) - F_Y(p|}{\max\{|F_X(q),F_Y(p)|\}+\epsilon}
 \end{equation}
 
Finally, a final score ${S_Y}$ for the model in evaluation is estimated through error pooling across all points, using: $S_Y = \frac{1}{N_p}\sum_{p=1}^{N_p}S_y(p)^k$
 
The implementation available online\footnote{https://github.com/mmspg/pointssim} was used. Color attributes were considered, as they led to the best results in the metric paper ~\cite{AlexiouPointSSIM}. Geometry attributes extracted from the metric were considered as well. The normal information attributed was tested but did not present any notable results. The variance $(\sigma^2)$ was chosen as a dispersion statistic, as recommended by the authors~\cite{AlexiouPointSSIM}.

\subsection{Point Cloud Quality Metric - PCQM~\cite{MeynetPCQM}}
PCQM~\cite{MeynetPCQM} implements a data-driven approach. The final quality score is computed as a linear combination of an optimal subject of the computed features. A correspondence is established between the distorted point cloud \textit{D} and the reference cloud \textit{R}, with a neighborhood defined for each point. This correspondence is obtained by fitting a quadric surface to a set of nearest neighbors $p_i^D\in D$ of $\textit{p}\in R$, considering a spherical neighborhood, which is based on the geometric correspondence between \textit{R} and \textit{D}. For each point in the computed quadric surface, the color of the nearest neighbor in \textit{D} is assigned.

The metric extracts eight features in total, defined as follows:
\begin{equation}\label{eq:f1}
\centering
    \text{Curvature comparison} \qquad f_1^{p} =\frac{\| \mu_p^{\rho} - \mu_{\hat{p}}^{\rho} \|}{\max (\mu_p^{\rho},\mu_{\hat{p}}^{\rho})+k_1}
\end{equation}

\begin{equation}\label{eq:f2}
\centering
    \text{Curvature contrast} \qquad f_2^{p} =\frac{\| \sigma_p^{\rho} - \sigma_{\hat{p}}^{\rho} \|}{\max (\sigma_p^{\rho},\sigma_{\hat{p}}^{\rho})+k_2} 
\end{equation}

\begin{equation}\label{eq:f3}
    \text{Curvature structure} \qquad f_3^{p} =\frac{\| \sigma_p^{\rho}\sigma_{\hat{p}}^{\rho} - \sigma_{p\hat{p}}^{\rho} \|}{\sigma_p^{\rho}\sigma_{\hat{p}}^{\rho}+k_3} 
\end{equation}

\begin{equation}\label{eq:f4}
    \text{Lightness comparison} \qquad f_4^{p}=\frac{1}{k_4 \cdot (\mu_p^{L}-\mu_{\hat{p}}^{L})^2+1} 
\end{equation}

\begin{equation}\label{eq:f5}
    \text{Lightness contrast} \qquad f_5^{p}=\frac{2\sigma_p^{L}\sigma_{\hat{p}}^{L}+k_5}{{\sigma_p^{L}}^2 + {\sigma_{\hat{p}}^{L}}^2+k_5} 
\end{equation}

\begin{equation}\label{eq:f6}
    \text{Lightness structure} \qquad f_6^{p}=\frac{\sigma_{p\hat{p}}^{L}+k_6}{\sigma_p^{L}\sigma_{\hat{p}}^{L}+k_6}
\end{equation}

 \begin{equation}\label{eq:f7}
    \text{Chroma comparison } \qquad  f_7^{p}=\frac{1}{k_7 \cdot (\mu_p^{C}-\mu_{\hat{p}}^{C})^2+1}
\end{equation}

\begin{equation}\label{eq:f8}
    \text{Hue comparison} \qquad f_8^{p}=\frac{1}{k_8 \cdot {\overline{\Delta H_{p\hat{p}}}^2+1}}
\end{equation}

The geometry features $f_1$, $f_2$ and $f_3$ are based on the curvatures ($\rho$). In equations \ref{eq:f1} to \ref{eq:f3}, $\mu_p^{\rho}$ and $\mu_{\hat{p}}^{\rho}$ represent Gaussian-weighted averages of curvature over the computed neighborhoods. Furthermore, $\sigma_p^{\rho}$, $\sigma_{\hat{p}}^{\rho}$ and $\sigma_p^{\rho}\sigma_{\hat{p}}^{\rho}$ are defined as the standard deviations and covariance of curvature over the aforementioned neighborhoods. The point cloud RGB values are converted to the perceptual color space LAB2000HL for the computation of luminance ($f_4$, $f_5$, $f_6$) and chroma features ($f_7$, $f_8$). In equations \ref{eq:f4} to \ref{eq:f6}, $ \mu_p^{L}$ and $\mu_{\hat{p}}^{L}$ represent the Gaussian-weighted averages of luminance over the computed neighborhoods, while $\sigma_p^{L}$, $\sigma_{\hat{p}}^{L}$ and $\sigma_p^{L}\sigma_{\hat{p}}^{L}$ are the standard deviations and covariance of luminance over the aforementioned neighborhood. Furthermore, $\mu_p^{C}$ and $\mu_{\hat{p}}^{C}$ are the Gaussian-weighted averages of chrominance in equation \ref{eq:f7}. Finally, in equation \ref{eq:f8}, $\Delta H_{p\hat{p}}=\sqrt{(a_p-a_{\hat{p}})^2+(b_p-b_{\hat{p}})^2 + (c_p-c_{\hat{p}})^2}$, in which $a_p$ and $b_p$ are two chromatic features and $c_p=\sqrt{a_p^2+b_p^2}$~\cite{MeynetPCQM}. $\overline{\Delta H_{p\hat{p}}}$ is the Gaussian-weighed average over $N(p,h)$~\cite{MeynetPCQM}.

The features are computed locally for each point \textit{p} between $[0,1]$, yielding $f_i^p$. The global features $f_i$ are then obtained by average pooling, using:  $f_i = \frac{1}{|R|} \sum_{p \in R} {f_i^p}$.

Finally, the perceptual quality score is given by a linear combination of the global using$ \mbox{PCQM} = \sum_{i \in S} {w_if_i}$,
where $S$ is the set of indices of the global features, while $f_i$ and $w_i$ refer to the $i^{th}$ global feature and its associated weight, respectively.

After optimization using logistic regression, the authors determined that the best features are $f_3$, $f_4$ and $f_6$, and the final recommended weights are  $PCQM_{rec}=0.18f_3+0.44f_4+0.38f_6$


\subsection{Multi Scale Graph Similarity - MS-GraphSIM~\cite{MSGraphSIM}}
MS-GraphSIM~\cite{MSGraphSIM} extends the GraphSIM metric~\cite{QiYangGraphSIM2022} by computing its graph-based features at different scales to better represent the characteristics of the Human Visualization System (HVS).

The metric starts by extracting a set of keypoints, $\overrightarrow{s}$, obtained by resampling the reference point cloud ($\overrightarrow{P}_r$) geometry using a high-pass graph filter. The resulting point cloud, $\overrightarrow{P}_s$, shows high spatial-frequency regions like edges or contours.
For both the reference and distorted ($\overrightarrow{P}_d$) point clouds, local graphs are constructed with each obtained keypoint $\overrightarrow{s}_k$ as its center. The neighbors of $\overrightarrow{s}_k$ are clustered using the Euclidean distance of the geometry components of corresponding points in $\overrightarrow{P}_r$ and $\overrightarrow{P}_d$.
After constructing a local graph, the color information of the neighbors belonging to that graph is set as a graph signal and passed through a low-pass graph filter, amplifying the low frequencies and attenuating the high frequencies. The neighborhood undergoes down-sampling using systematic sampling~\cite{MSGraphSIM}. The sampled points are moved towards the centroid of the point cloud bounding box. For each scale, three similarity measures are computed based on color gradient features, notably the gradient mass ($m_g$), the gradient mean ($\mu_g$) and the gradient variance $\sigma^2_g$ and covariance ($c_g$). These features reflect the spatial variation of point density, the distortion of the statistical characteristics of the signals, and the spatial disturbance of the points, respectively, and are defined as follows:

\begin{equation}\label{eq:gradMass}
    m_g=\sum_{\overrightarrow{X}_j \in \mathcal{N}_k}\sqrt{W_{\overrightarrow{X}_j, \overrightarrow{s}_k}}[f(\overrightarrow{X}_j) - f(\overrightarrow{s}_k)]
\end{equation}

\begin{equation}\label{eq:gradMean}
    \mu_g=\frac{1}{N}(\nabla_{\overrightarrow{s}_k}f)
\end{equation}

\begin{equation}\label{eq:var}
    \sigma^2_g=\frac{\sum(g_j - \overline{g})}{N}
\end{equation}

\begin{equation} \label{eq:cov}
    c_g= E[\overrightarrow{g}_{\overrightarrow{s}_k} \cdot \overline{g}_{\overrightarrow{s}_k}] - E[\overrightarrow{g}_{\overrightarrow{s}_k}] \cdot E[\overline{g}_{\overrightarrow{s}_k}]
\end{equation}

In equation (\ref{eq:gradMass}), $W_{\overrightarrow{X}_j,\overrightarrow{s}_k}$ represents the euclidean-based graph weight,$\overrightarrow{X}_j$ represents the color attributes of the point cloud, and $f(\overrightarrow{X}_j) - f(\overrightarrow{s}_k)$ is the attribute gradient. $\mathcal{N}_k$ is the set of points effectively connected to $\overrightarrow{s}_k$. In equation (\ref{eq:gradMean}), $N$ represents the number of points in $\mathcal{N}_k$.
In equation (\ref{eq:var}), $g_j$ represents the \textit{j}-th element in $\overrightarrow{g}_{\overrightarrow{s}_k}$ and $\overline{g}$ is the weighted average gradient of $\overrightarrow{g}_{\overrightarrow{s}_k}$.
Finally, in equation (\ref{eq:cov}), $\overrightarrow{g}_{\overrightarrow{s}_k}$ and $\overline{g}_{\overrightarrow{s}_k}$ represent the weighted gradient distribution of both the reference and impaired point clouds.

Finally, the three similarity measures, i.e., $SIM_{m_g}$, $SIM_{\mu_g}$, and $SIM_{c_g}$, are obtained as follows:
\begin{equation}\label{SSIMmg}
    SIM_{m_g}  = \frac{2m_g^r \cdot m_g^d + T_0}{(m_g^r)^2 + (m_g^d)^2 + T_0}
\end{equation}
\begin{equation}\label{SSIMmug}
    SIM_{\mu_g} = \frac{2\mu_g^r \cdot \mu_g^d + T_1}{(\mu_g^r)^2 + (\mu_g^d)^2 + T_1}
\end{equation}
\begin{equation}\label{SSIMcov}
    SIM_{c_g}  = \frac{c_g d + T_2}{\sigma_g^r \cdot \sigma_g^d + T_2}
\end{equation}

\begin{table*}[b]
    \centering
    \large
    \caption{PCC/SROCC  between the considered metrics for the BASICS database.}
    \label{tab:BascisComparison}
    \resizebox{\linewidth}{!}{%
    \begin{tabular}{|c|c|c|c|c|c|c|c|c|c|c|c|c|c|c}
    \hline
       vs  & PSNR MSE D1 & PSNR MSE D2 & PSNR MSE YUV & \makecell[c]{PointSSIM \\ Geometry Features} & \makecell[c]{PointSSIM \\ Luminance Features} & GraphSIM & MS-GraphSIM & PCQM \\ \hline \hline
       MOS & 0.841 / 0.785 & 0.897 / \textbf{0.847} & 0.603 / 0.602 & 0.782 / 0.750 & 0.754 / 0.718 & \textbf{0.913} / 0.846 & 0.911 / 0.834 & \textbf{0.913} / 0.843 \\ \hline
        PCQM  & 0.803 / 0.742 & 0.814 / 0.815 & 0.687 / 0.674 & 0.808 / 0.761 & 0.730 / 0.714 & 0.945 / 0.940 & 0.912 / 0.894  \\ \cline{1-8}
        MS-GraphSIM & 0.852 / 0.863 & 0.857 / 0.899 & 0.627 / 0.621 & 0.877 / 0.874 & 0.847 / 0.830 & 0.978 / 0.972     \\ \cline{1-7}
        GraphSIM &  0.834 / 0.812 & 0.835 / 0.853 & 0.617 / 0.617 & 0.868 / 0.832 & 0.809 / 0.789         \\ \cline{1-6}
        \makecell[c]{PointSSIM \\ Luminance Features}  & 0.707 / 0.756 & 0.749 / 0.783 & 0.416 / 0.351 & 0.614 / 0.685    \\ \cline{1-5}
        \makecell[c]{PointSSIM \\ Geometry Features}  & 0.869 / 0.880 & 0.827 / 0.869 & 0.577 / 0.576  \\ \cline{1-4}
        PSNR MSE YUV & 0.527 / 0.554 & 0.558 / 0.628     \\ \cline{1-3}
        PSNR MSE D2 & 0.956 / 0.935      \\ \cline{1-2}
    \end{tabular}%
    }
\end{table*}

\noindent where $T_0$, $T_1$ and $T_2$ are non-zero constants defined to prevent numerical instability, set to 0.001..
The overall similarity is given by $S_{\overrightarrow{s}_k,C} = SIM_{m_g} \cdot SIM_{\mu_g} \cdot SIM_{c_g}$, which is then pooled across all color channels, using:  $S_{\overrightarrow{s}_k}=\frac{1}{\gamma}\sum_C \gamma C \cdot |S_{\overrightarrow{s}_k,C}|$, where $\gamma C$ is the pooling factor that reflects the importance of each color channel in the visual perception. The point cloud RGB components are decomposed to the Color Gaussian model~\cite{QiYangGraphSIM2022}. This results in a luminance component ($\hat{E}$) and two chrominance components ($\hat{E}_\lambda$ and $\hat{E}_{\lambda\lambda}$). As such, the authors set the pooling factor as $\hat{E}: \hat{E}_\lambda:\hat{E}_{\lambda\lambda}=6:1:1$, as in the overall PSNR calculation of YUV~\cite{PSNRYUV}.
The final overall similarity score is obtained by averaging across the total number of keypoints.
Finally, for each scale, the overall quality scores are aggregated using the following pooling operation:
\begin{equation}
    Q_{overall} =\frac{\sum^{M}_{i=0}w_iS_{\overrightarrow{s}_k}}{\sum^{M}_{i=0}w_i}
\end{equation}
\noindent where $M$ represents the different scales and $w_i$ denotes the weighted factor for the different scales.

\section{Initial Approach}
Initially, the Pearson Correlation Coefficient (PCC) and the Spearman rank-order correlation coefficient (SROCC) between the metrics results and the Mean Opinion Score (MOS) are computed after logistic fitting (described in Section \ref{sec:RFEAnalysis}), for the BASICS database. The results are shown in Table \ref{tab:BascisComparison}.

Overall, metrics that consider both the geometry and the color attributes achieve the best performance, while metrics that consider only the color attributes perform worst.

This can be observed with metrics like GraphSIM, MS-GraphSIM, and PCQM that achieve PCC values above 0.9.
Metrics that only use geometry achieve similar PCC values, but PSNR MSE D2 presents a higher SROCC, indicating it provides better monotonic behavior. In this set of metrics, PointSSIM has the lowest values.
The luminance features of PointSSIM show the highest correlation values among the metrics that only use color attributes.
\begin{figure}[!t]
    \centering
    \subfigure[\textit{GraphSIM vs MS-GraphSIM}]{\includegraphics[width=0.44\linewidth]{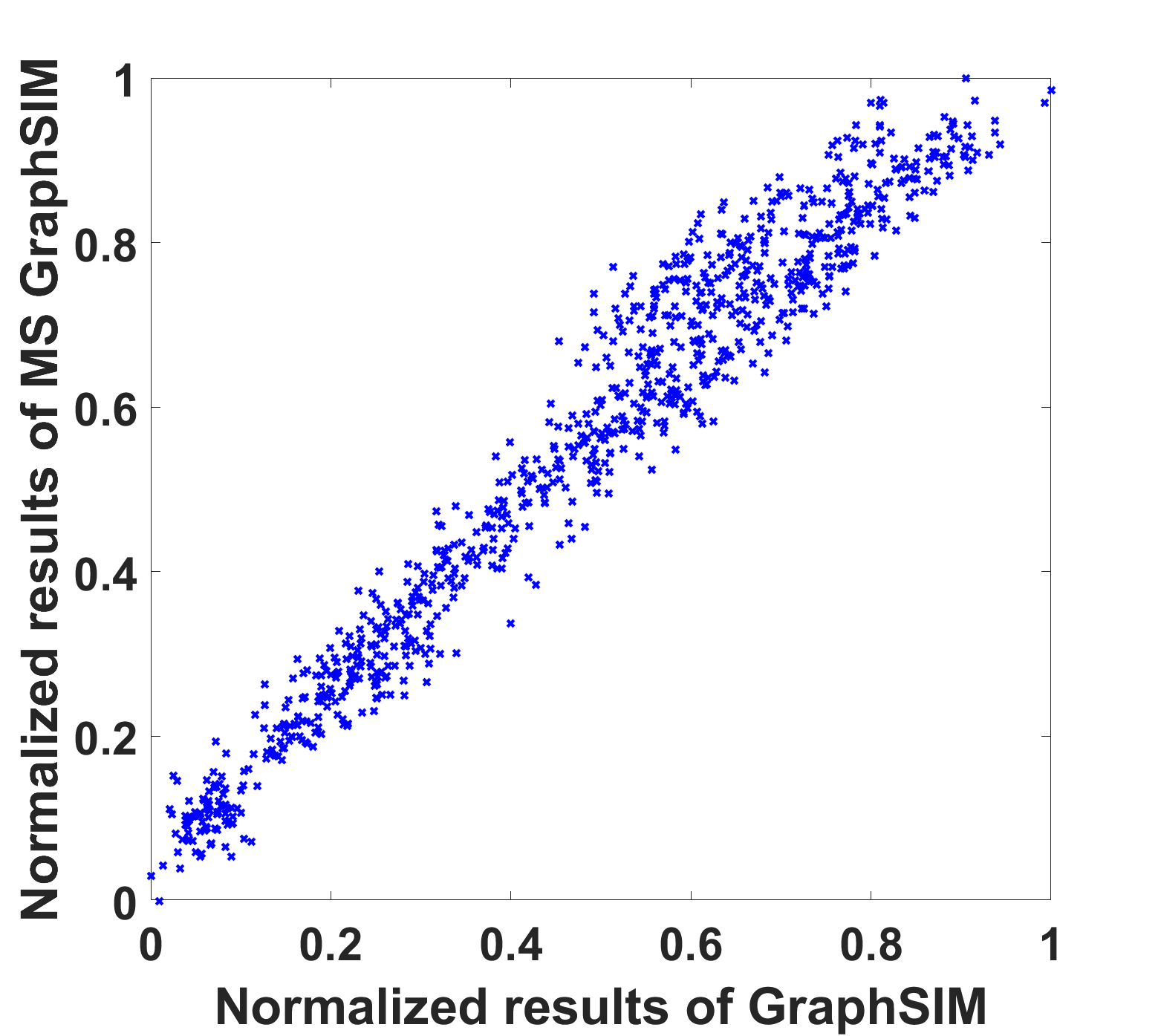}}\hfill
    \subfigure[\textit{PSNR MSE D2 vs PSNR MSE D1}]{\includegraphics[width=0.44\linewidth]{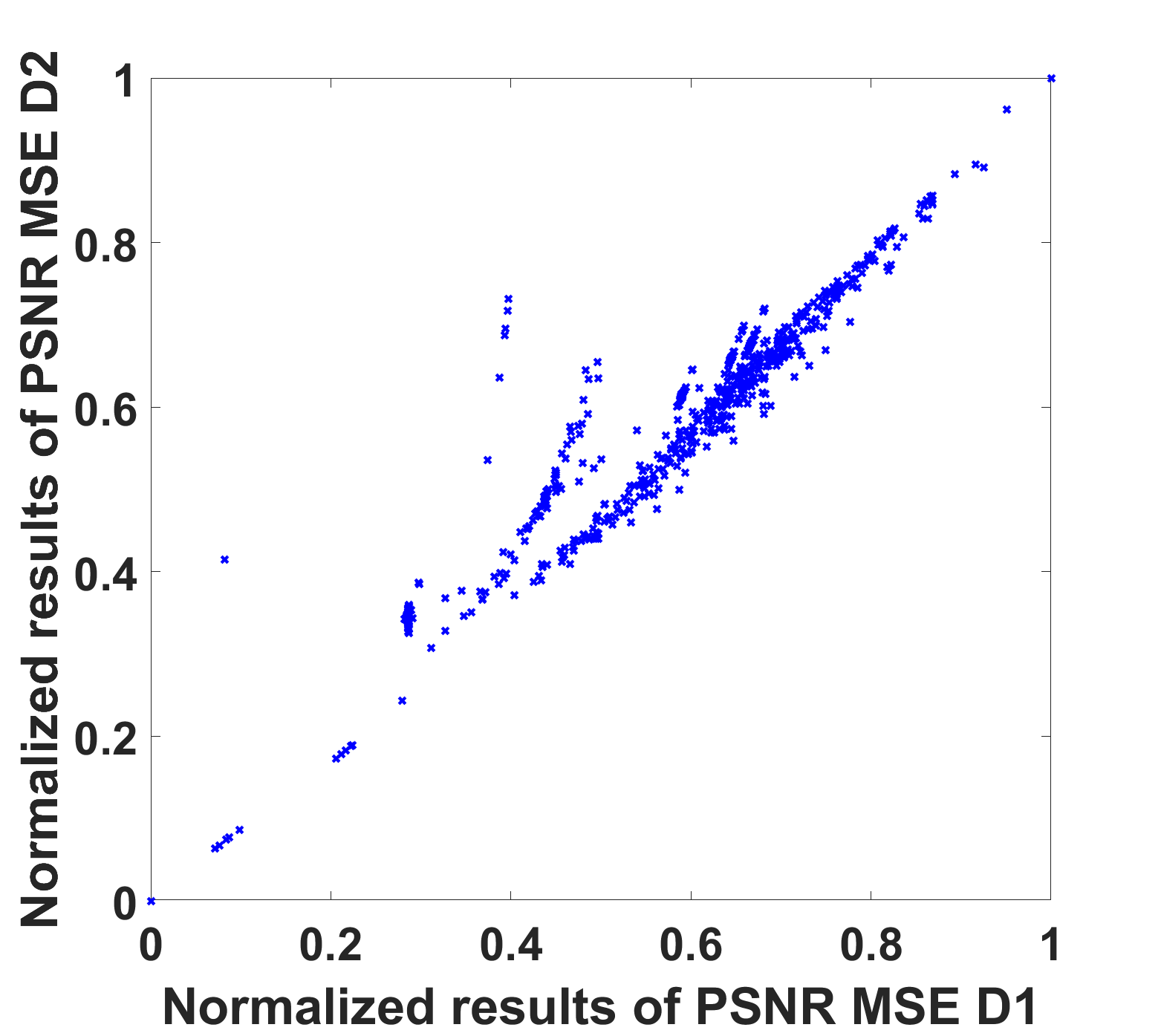}} \hfill \\
    \subfigure[\textit{PCQM vs MS-GraphSIM}]{\includegraphics[width=0.44\linewidth]{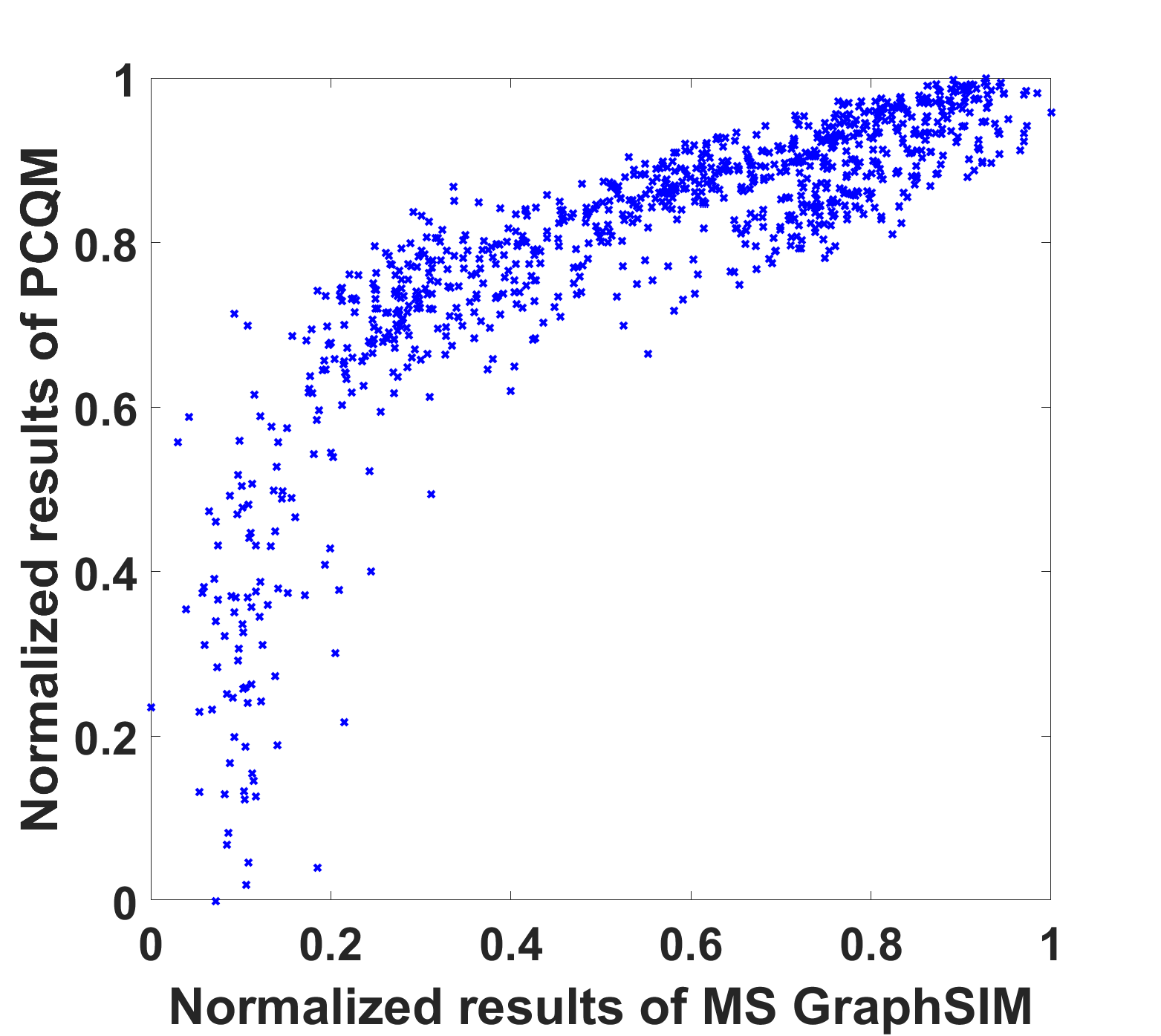}} \hfill
    \subfigure[\textit{PSNR MSE YUV vs MS-GraphSIM}]{\includegraphics[width=0.44\linewidth]{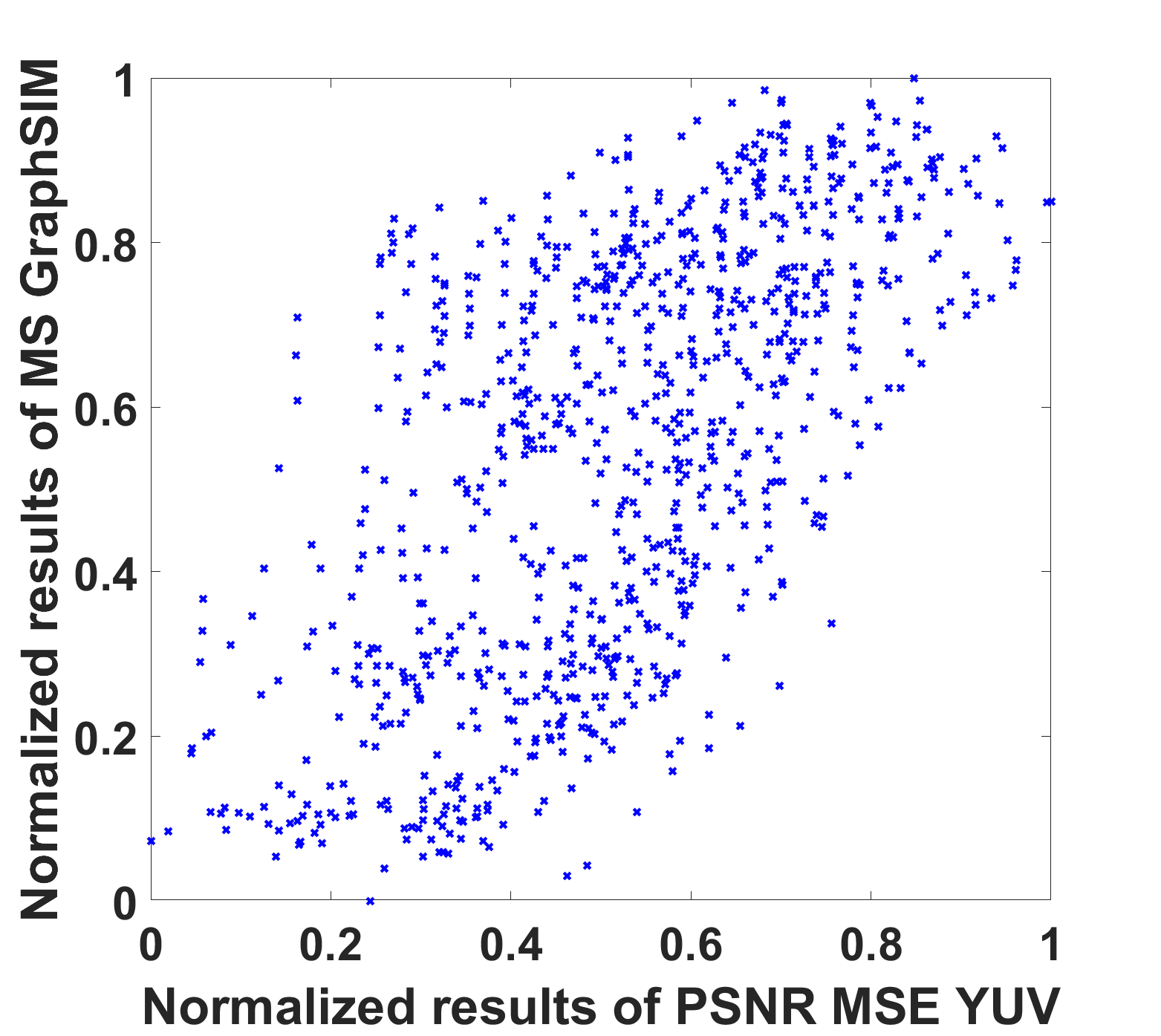}} \hfill\\

    \caption{Normalized results of some metrics for the BASICS Database.}
    \label{fig:BasicsComparison}
\end{figure}
Table \ref{tab:BascisComparison} also shows the PCC/SROCC between the different metrics also after logistic fitting for the BASICS database.
Some metrics exhibit high correlations between them, such as PSNR MSE D1 vs. PSNR MSE D2, GraphSIM vs. MS-GraphSIM, PCQM vs. MS-GraphSIM, and PCQM vs. GraphSIM. The PSNR MSE D1 and PSNR MSE D2 metrics consider very similar information.

Furthermore, PSNR MSE YUV achieved small correlation values with the other metrics that only considers color attributes, notably the PointSSIM Luminance Features. 

Figure \ref{fig:BasicsComparison} shows the normalized values between some metrics with high and low correlations. It can be observed that for metrics with high correlation values, such as GraphSIM vs MS-GraphSIM and PSNR MSE D2 vs PSNR MSE D1, the plot is not sparse, but as the correlation decreases, the plot becomes more scattered.

For simplicity reasons, in the study, only MS-GraphSIM is referred, since GraphSIM is the first scale of MS-GraphSIM.  
Although these metrics have high correlation (PCC/SROCC), it is considered that the representation of quality provided by each metric is different.  
In that sense, it is possible to explore the different representations provided by the different metrics to improve the representation of the subjective quality.

\section{Regression methods and Feature Analysis} \label{sec:FeatureAnalysis}
The considered metrics allow studying a wide range of features, taking in consideration geometry, luminance, chrominance and normal information, as shown in table \ref{tab:MetricChar}. Furthermore, the table also shows the Pearson (PCC) and Spearman (SROCC) correlations obtained for each considered feature. In total, 23 features are considered. $SIM_{m_g}$ achieved the highest correlation values, followed by PSNR MSE D2 and $f4$. The performance of PSNR MSE D2 is most likely due to the fact that the learning-based codec present in the database was optimized using that feature.

In the following section, a description of the regression methods and Feature Analysis methodology is provided.
\begin{table*}
    \centering
    \caption{Information considered by each feature from the full reference metrics considered in this study.}
    \label{tab:MetricChar}
    \resizebox{\linewidth}{!}{%
    \begin{tabular}{|l|c|c|c|c|c|c|c|c|}
    \hline
         Metric   & Geometry & Normal Information & Luminance & Chrominance & \textbf{PCC} & \textbf{SROCC}  \\ \hline
         PSNR MSE D2 (Geometry)&  \checkmark & \checkmark & x & x  & 0.897 & \textit{0.847} \\ \hline 
         PSNR MSE Y (Luminance) &  x & x & \checkmark & x & 0.557 & 0.554\\ \hline
         PSNR MSE U (Chrominance) &  x & x & x & \checkmark & 0.561  & 0.54 \\ \hline 
         PSNR MSE V (Chrominance) &  x & x & x & \checkmark & 0.613  & 0.599 \\ \hline 
         PointSSIM Luminance Features  & x & x & \checkmark & x & 0.754  & 0.718\\ \hline 
         PointSSIM Geometry Features  & \checkmark & x & x & x & 0.782  & 0.750\\ \hline
         PCQM $f1$ (Geometry) &  \checkmark & x & x  & x & 0.858 & 0.821\\ \hline
         PCQM $f2$ (Geometry) &  \checkmark & x &  x & x & 0.868 & 0.829\\ \hline
         PCQM $f3$ (Geometry) &  \checkmark & x &  x & x & 0.759 & 0.682\\ \hline
         PCQM $f4$ (Luminance) &  x & x & \checkmark & x & \textit{0.899} & 0.818\\ \hline
         PCQM $f5$ (Luminance) &  x & x & \checkmark & x & 0.838 & 0.723\\ \hline
         PCQM $f6$ (Luminance) &  x & x & \checkmark & x & 0.840 & 0.790\\ \hline
         PCQM $f7$ (Chroma) &  x & x & x & \checkmark & 0.808 & 0.700 \\ \hline
         PCQM $f8$ (Hue) &  x & x  & x & \checkmark & 0.589 & 0.550\\ \hline
         MS-GraphSIM (Geometry + Luminance) $SIM_{m_g}$ &  \checkmark & x & \checkmark & x & \textbf{0.909}  & \textbf{0.848}\\\hline
         MS-GraphSIM (Geometry + Chroma) $SIM_{\mu_g}$ &  \checkmark & x & x & \checkmark & 0.889 & 0.825\\ \hline
         MS-GraphSIM (Geometry + Chroma) $SIM_{c_g}$ &  \checkmark & x & x & \checkmark & 0.887 & 0.810\\ \hline
         MS-GraphSIM (Geometry + Luminance) $SIM_{m_g}$ Scale 1 &  \checkmark & x & \checkmark & x & 0.786 & 0.698\\\hline
        MS-GraphSIM (Geometry + Chroma) $SIM_{\mu_g}$ Scale 1&  \checkmark & x & x & \checkmark & 0.776 & 0.702\\ \hline
         MS-GraphSIM (Geometry + Chroma) $SIM_{c_g}$ Scale 1&  \checkmark & x & x & \checkmark & 0.772 &0.699\\ \hline
         MS-GraphSIM (Geometry + Luminance) $SIM_{m_g}$ Scale 2&  \checkmark & x &\checkmark & x & 0.857 & 0.767\\\hline
         MS-GraphSIM (Geometry + Chroma) $SIM_{\mu_g}$ Scale 2&  \checkmark & x & x & \checkmark & 0.855 & 0.742\\ \hline
        MS-GraphSIM (Geometry + Chroma) $SIM_{c_g}$ Scale 2&  \checkmark & x & x & \checkmark &0.833 & 0.740\\ \hline
    \end{tabular}%
    }
\end{table*}

\subsection{Regression method}
For regression methods, an SVR~\cite{SVR2015} and Ridge Regression~\cite{RidgeRegression} are considered. 
Support Vector Regression (SVR) aims to find a hyperplane that best fits the data while allowing for some margin of error, focusing on points within a tolerance level. The Ridge Regression adds a penalty to the size of coefficients in a linear model to prevent overfitting by discouraging large coefficients.
The alternative of using a deep learning-based classification was not selected because the amount of training data is not large enough to provide a reliable metric.
Also, the use of data augmentation is not advisable, as the changes created by these models are likely to cause changes in subjective quality, which will lead to unreliable training data.

For data pre-processing, a Min-Max normalization method was employed to scale the extracted features. Finally, the Python sklearn package\footnote{https://scikit-learn.org/stable/} was considered to implement a kernel SVR model with a radial basis function (RBF), and the RR model using $\alpha = 1$, after hyperparameter optimization.

\subsection{Dataset For the Feature Study}\label{sec:Datasets}

\begin{figure}
    \centering
    \subfigure[\textit{p13} (Basics)]{\includegraphics[width=0.19\linewidth]{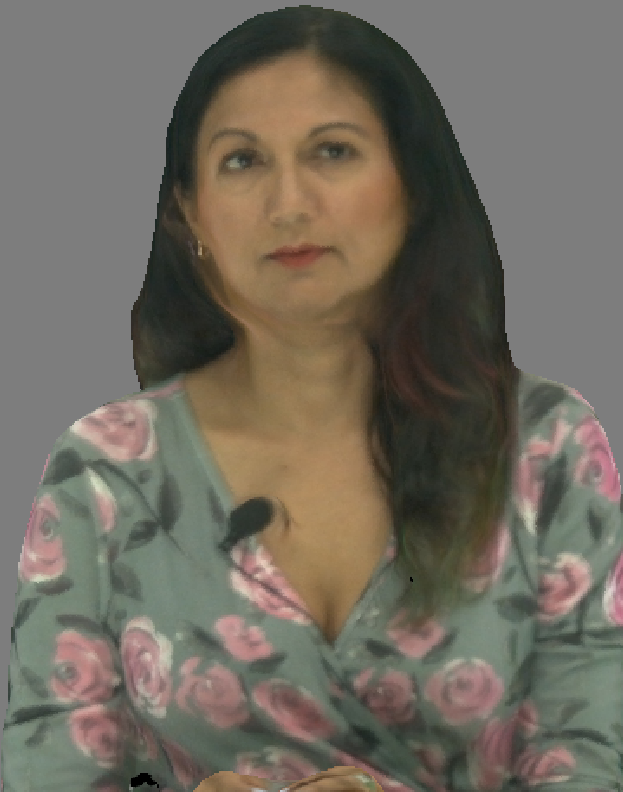}}
    \subfigure[\textit{p13} V-PCC]{\includegraphics[width=0.19\linewidth]{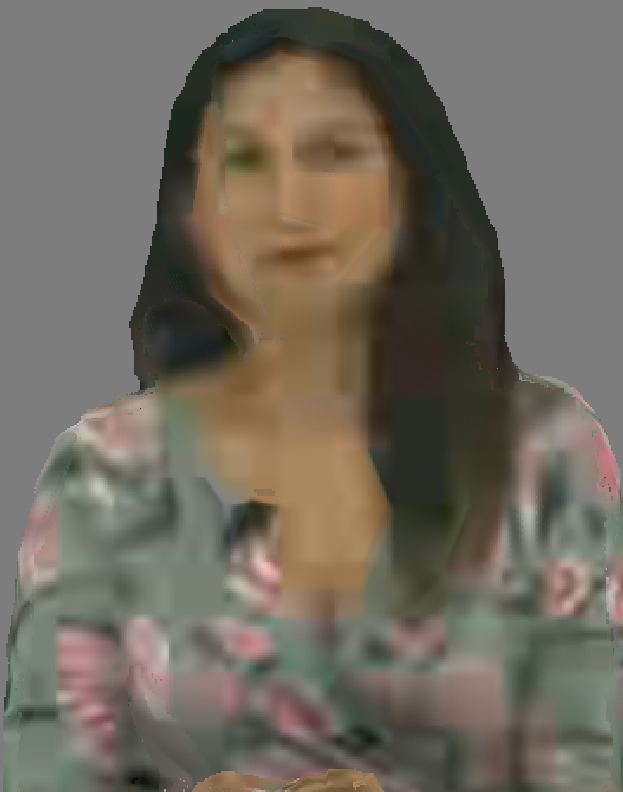}}
    \subfigure[\textit{p13} G-PCC Predlift]{\includegraphics[width=0.19\linewidth]{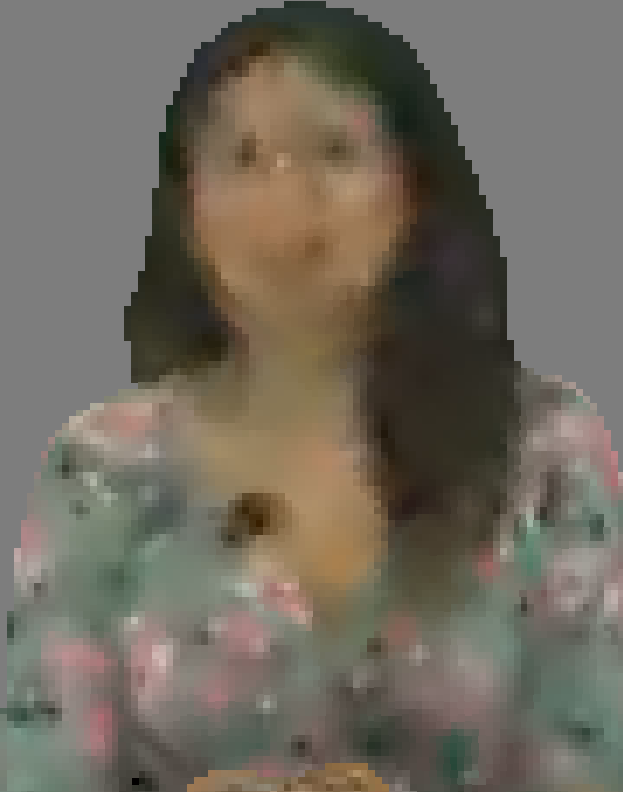}}
    \subfigure[\textit{p13} G-PCC RAHT]{\includegraphics[width=0.19\linewidth]{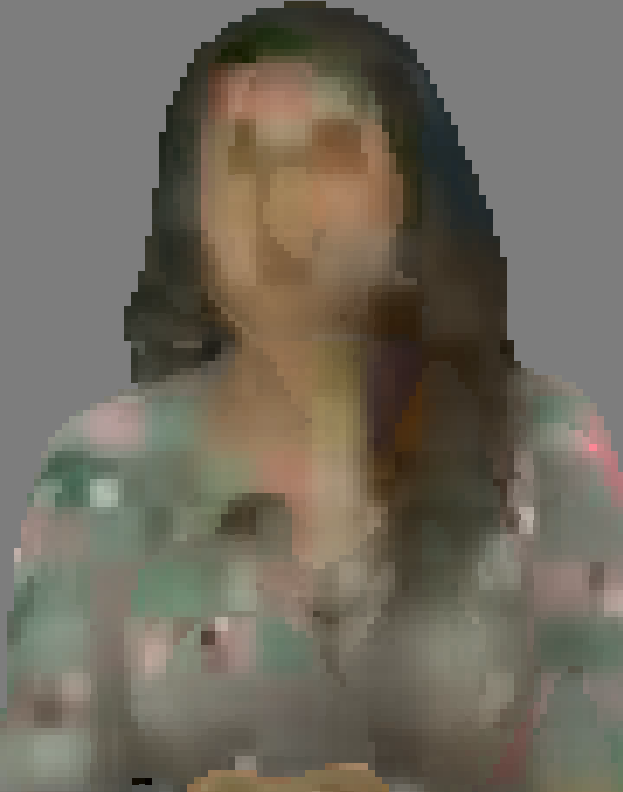}} 
    \subfigure[\textit{p13} GeoCNN]{\includegraphics[width=0.19\linewidth]{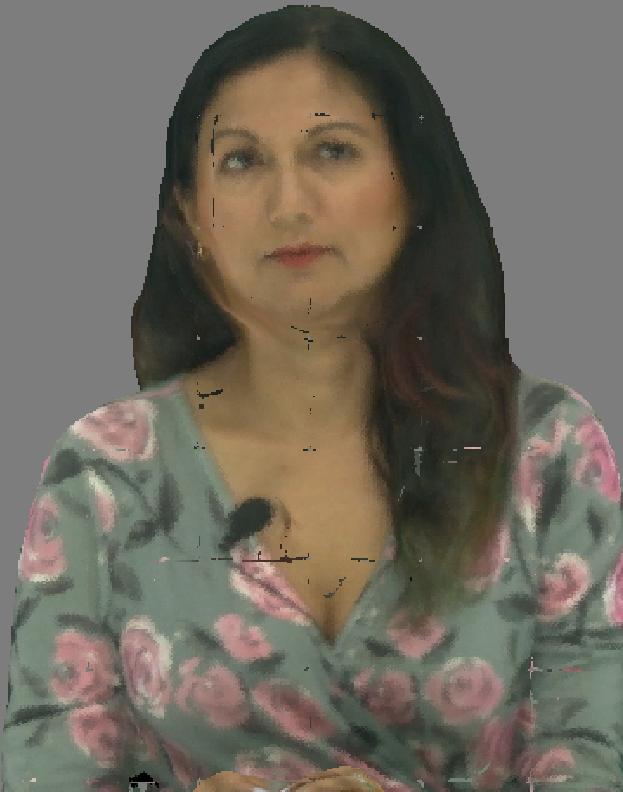}} \\
    \subfigure[\textit{p24}]{\includegraphics[width=0.19\linewidth]{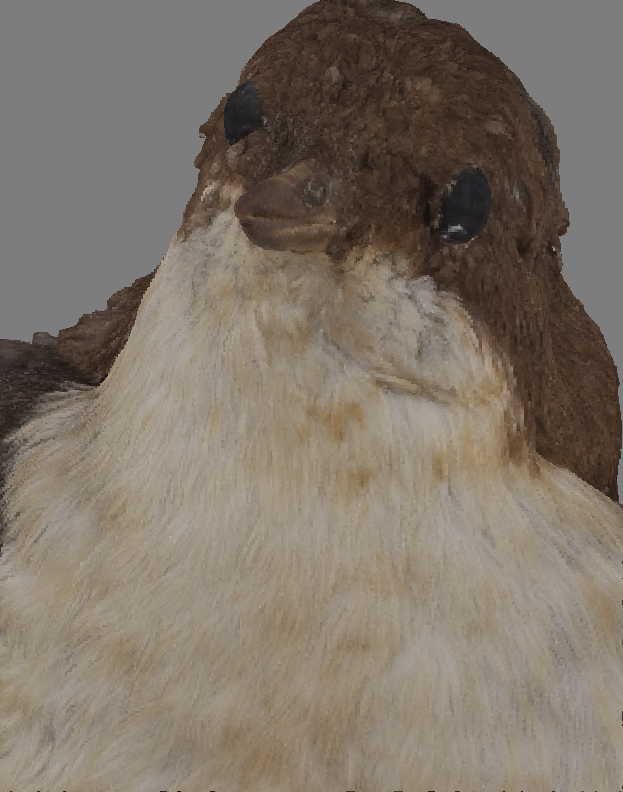}}
    \subfigure[\textit{p24} V-PCC]{\includegraphics[width=0.19\linewidth]{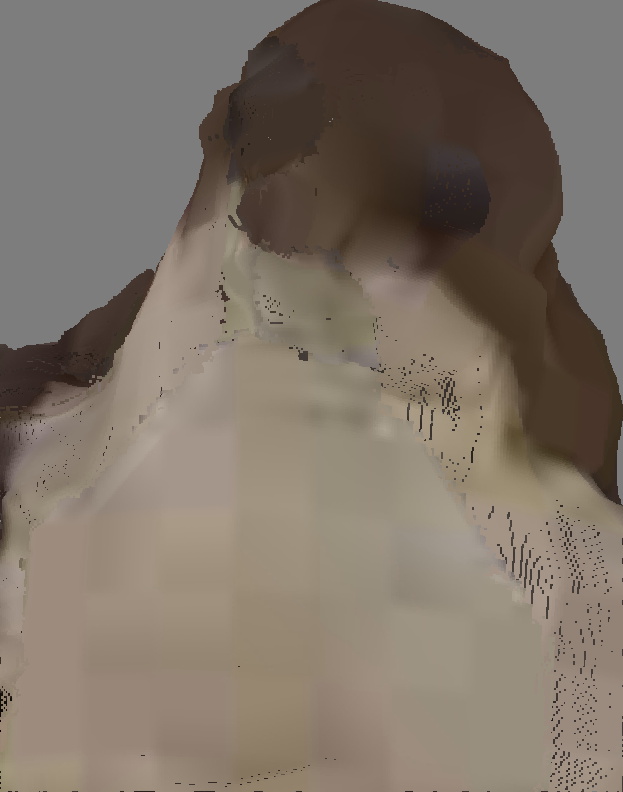}}
    \subfigure[\textit{p24} G-PCC Predlift]{\includegraphics[width=0.19\linewidth]{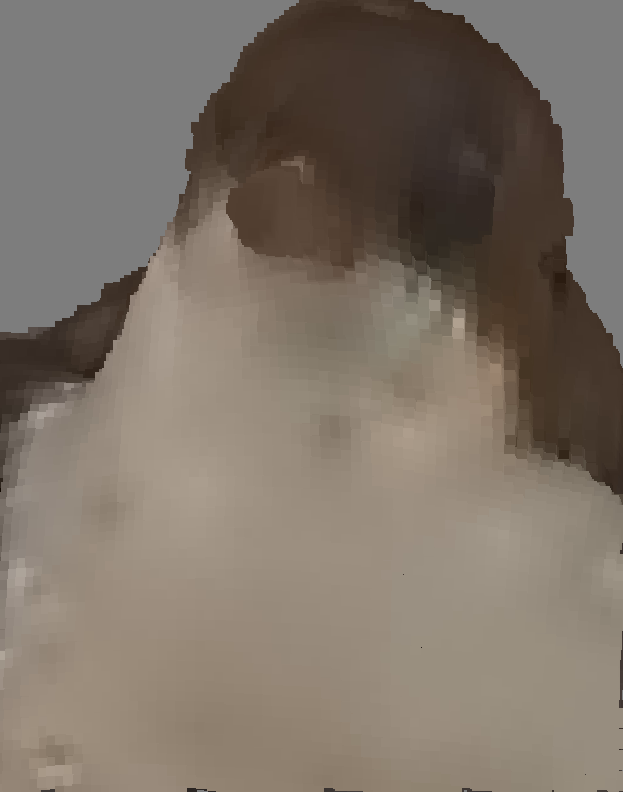}} 
    \subfigure[\textit{p24} G-PCC RAHT]{\includegraphics[width=0.19\linewidth]{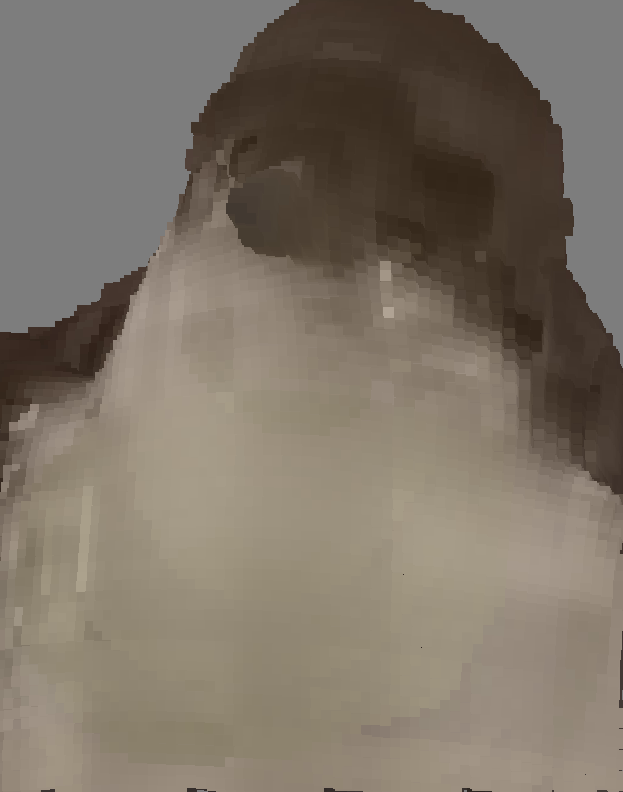}} 
    \subfigure[\textit{p24} GeoCNN]{\includegraphics[width=0.19\linewidth]{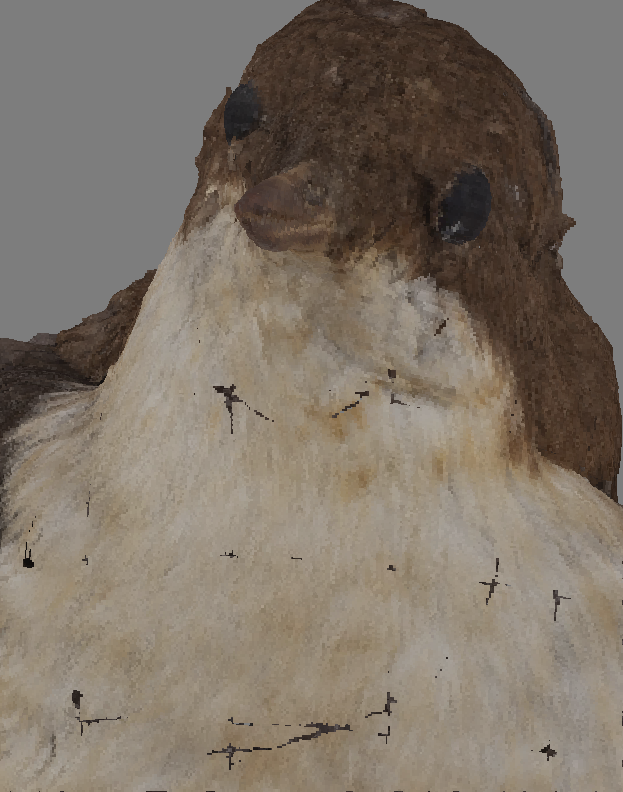}} \\
    \subfigure[\textit{p72}]{\includegraphics[width=0.19\linewidth]{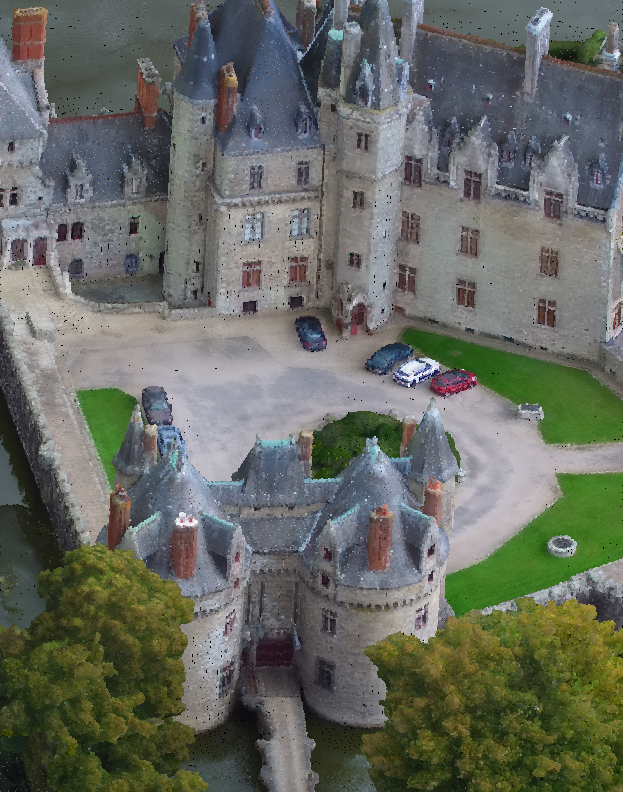}}
    \subfigure[\textit{p72} V-PCC]{\includegraphics[width=0.19\linewidth]{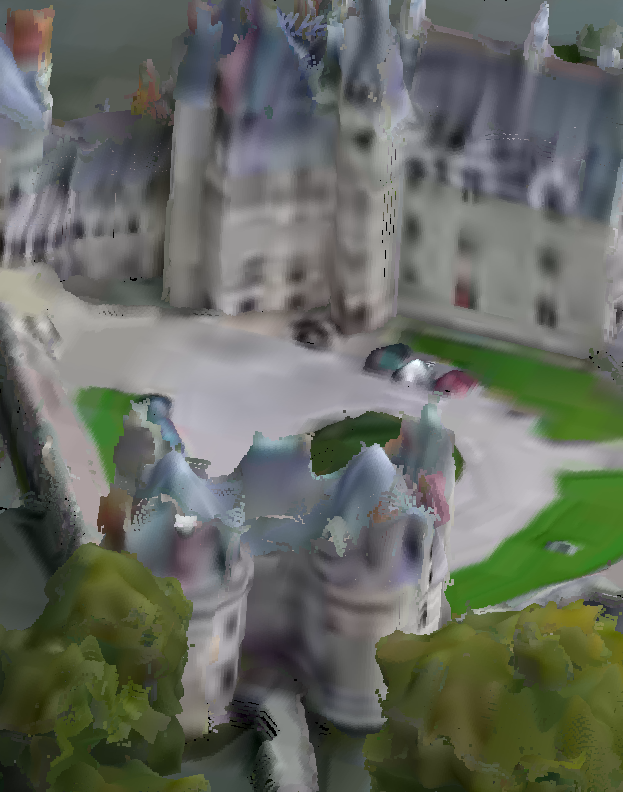}}
    \subfigure[\textit{p72} G-PCC Predlift]{\includegraphics[width=0.19\linewidth]{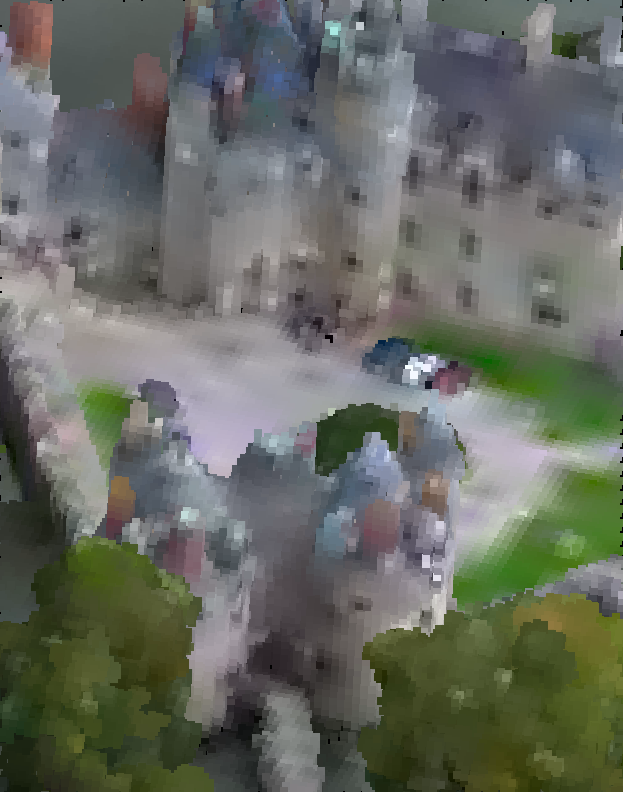}}
    \subfigure[\textit{p72} G-PCC RAHT]{\includegraphics[width=0.19\linewidth]{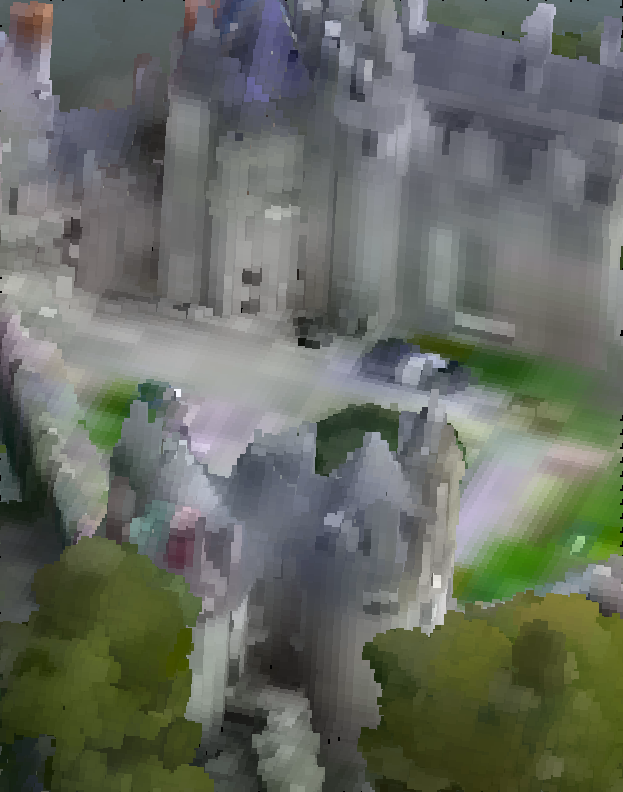}}
    \subfigure[\textit{p72} GeoCNN]{\includegraphics[width=0.19\linewidth]{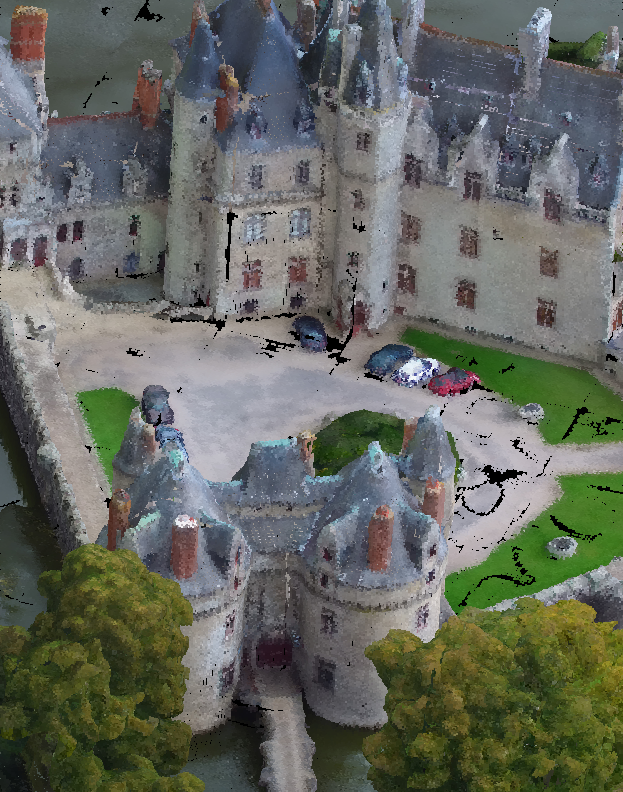}}
    \caption{Examples of point clouds in the BASICS Database. The first column shows the reference point cloud, and the remaining ones depicts the lowest rate that results from each codec.}
    \label{fig:Examples}
\end{figure}

The Broad Quality Assessment of Static Point Clouds in Compression Scenario (BASICS) training dataset~\cite{ak2023basics} was selected to study the contribution of each feature for the prediction of objective quality. This database contains 898 coded point clouds, with distortions introduced by different coding methods, notably the octree model of G-PCC~\cite{VPCCandGPCC}, using both the RAHT~\cite{Queiroz2016a} and Predlift~\cite{VPCCandGPCC} methods, the video-based codec V-PCC~\cite{VPCCandGPCC}, and the learning-based solution GeoCNN~\cite{quach2020improved}, from 45 reference point clouds.
Furthermore, the point clouds represent several scenarios. Figure \ref{fig:Examples} shows three examples of point clouds from BASICS, representing human content (\textit{p13}), a bird (\textit{p24}), and a landscape (\textit{p72}). The results obtained for the lowest coding rates are shown, so that the distortions typically created by these aforementioned coding solutions are well visible.

\begin{table*}[t]
\caption{Metric performance using ten-fold cross validation using the BASICS training dataset.}
\label{tab:crossValidation}  
\resizebox{\linewidth}{!}{%
\begin{tabular}{@{}|l|c|l|c|c|c|c|c|c|@{}}
\hline
 \textbf{\textit{\makecell[l]{Metric\\ Combination}}} & \textbf{\makecell[c]{Regression \\ Method}} & \textit{Features} & \textit{Type} & \textbf{\textit{PCC}} & \textbf{\textit{$\sigma_{PCC}$}} & \textbf{\textit{SROCC}}& \textbf{\textit{$\sigma_{SROCC}$}} \\ \hline
\hline
\makecell[l]{Model 1 \\(8 features)} & SVR & \makecell[l]{PCQM ($f2$, $f4$, $f5$, $f6$), MS-GraphSIM ($SIM_{m_g}$ Scale 0, \\ $SIM_{\mu_g}$ Scale 0, $SIM_{c_g}$ Scale 0) PSNR MSE D2} & $FR,\,GEO+COL$& 0.939 & 0.018 & 0.873 & 0.026 \\ \hline  
\makecell[l]{Model 2 \\(10 features)} &  SVR & \makecell[l]{PCQM ($f2$, $f4$, $f5$, $f6$, $f7$), MS-GraphSIM ($SIM_{m_g}$ Scale 0, $SIM_{c_g}$ Scale 0),\\ PSNR MSE D2, PointSSIM Geometry and Luminance features} & $FR,\,GEO+COL$ & 0.939 & 0.022 & 0.886 & 0.043  \\ \hline
\makecell[l]{Model 3 \\(14 features)} & SVR & \makecell[l]{PCQM ($f2$, $f4$, $f5$, $f7$, $f8$), MS-GraphSIM ($SIM_{m_g}$ Scale 0,\\ $SIM_{\mu_g}$ Scale 0, $SIM_{m_g}$ Scale 0, $SIM_{\mu_g}$ Scale 2, and $SIM_{c_g}$ Scale 2), \\PSNR MSE D2, PSNR MSE V, \\ PointSSIM Geometry and Luminance features,} & $FR,\,GEO+COL$ & \textit{0.949} & \textit{0.013}& \textit{0.892} &0.022 \\ \hline
\makecell[l]{Model 4 \\(4 features)}& SVR & PCQM ($f2$, $f4$, $f5$), MS-GraphSIM ($SIM_{m_g}$ Scale 0)  & $FR,\,GEO+LUM$& 0.944 & 0.022 & 0.878 & 0.03 \\ \hline
\makecell[l]{Model 5 \\(6 features)} & RR & \makecell[l]{PCQM ($f2$, $f4$, $f5$, $f7$), \\MS-GraphSIM ($SIM_{m_g}$ Scale 0), PSNR MSE D2} & $FR,\,GEO+LUM$ & 0.932 & 0.025 & 0.869 & 0.035 \\ \hline
\makecell[l]{Model 6 \\(11 features)}& RR &\makecell[l]{PCQM ($f2$, $f4$, $f5$, $f7$, $f8$), MS-GraphSIM ($SIM_{m_g}$ Scale 0,\\ $SIM_{c_g}$ Scale 0, $SIM_{m_g}$ Scale 2, $SIM_{c_g}$ Scale 2),\\ PSNR MSE D2, PointSSIM Geometry Features} & $FR,\,GEO+COL$  & 0.944 & 0.024 & 0.887 & \textit{0.016}  \\ \hline
\makecell[l]{Model 7\\ (15 features)} & RR &  \makecell[l]{PCQM ($f1$,$f2$,$f4$,$f5$,$f7$,$f8$), MS-GraphSIM ($SIM_{m_g}$ Scale 0,\\$SIM_{c_g}$ Scale 0, $SIM_{c_g}$ Scale 1, $SIM_{c_g}$Scale 2), \\PSNR MSE D2, PSNR MSE Y,\\ PSNR MSE U, PSNR MSE V,PointSSIM Geometry Features} &$FR,\,GEO+COL$ &\textbf{0.951} & \textbf{0.01} & \textbf{0.9} & \textbf{0.014}  \\ \hline
\makecell[l]{Model 8 \\(4 features)}& RR &PCQM ($f2$, $f4$, $f5$), MS-GraphSIM ($SIM_{m_g}$ Scale 0) & $FR,\,GEO+LUM$&0.927 & 0.027 &0.87 &0.04 \\ \hline
\end{tabular}%
}
\end{table*}

\begin{table*}
\caption{Metric performance using the BASICS validation dataset. The metric combination models are defined in table~\ref{tab:crossValidation}}.
\label{tab:Validation}  
{\small
\begin{tabular}{@{}|l|c|c|c|c|c|c|c|@{}}
\hline
\textbf{\textit{Metric}}&\textbf{\makecell[c]{Regression \\ Method}}& \textit{Type} & \textbf{\textit{PCC}} & \textbf{\textit{SROCC}} & \textbf{\textit{RMSE}} & \textbf{\textit{OR}} & \makecell[l]{Average\\ Time (s)} \\ \hline
Model 1 (8 features) & SVR &$FR,\,GEO+COL$ & 0.936 & \textbf{0.881} & 0.098 & 0.717 & 66.324 \\ \hline  
Model 2 (10 features)& SVR  & $FR,\,GEO+COL$ & 0.924 & 0.846 & 0.106 & 0.727 & 86.724 \\ \hline
Model 3 (14 features) & SVR & $FR,\,GEO+COL$ & 0.933 & 0.845 & 0.100 & 0.683 & 109.312  \\ \hline
Model 4 (4 features) & SVR & $FR,\,GEO+LUM$ & 0.937 & 0.840 & 0.097 & 0.707 & 43.736\\ \hline
Model 5 (6 features) &RR &$FR,\,GEO+LUM$ & \textbf{0.944} & 0.854 & \textbf{0.092} &0.670 & 66.32\\ \hline  
Model 6 (11 features)& RR  & $FR,\,GEO+COL$  & 0.936 & 0.828 & 0.099 & 0.697 & 76.62\\ \hline
Model 7 (15 features) & RR & $FR,\,GEO+COL$ & 0.938 & 0.840 & 0.097 & 0.690 & 99.208  \\ \hline
Model 8 (4 features) & RR & $FR,\,GEO+LUM$& 0.930 & 0.832 & 0.103 & 0.687 & 43.732 \\ \hline

\hline
PSNR MSE D1~\cite{Dtian} & - & $FR,\,GEO$ & 0.894 & 0.800 & 0.126 & 0.760 & 22.588 \\ \hline
PSNR MSE D2~\cite{Dtian} & - & $FR,\,GEO$ & 0.923 & 0.836 & 0.108 & 0.693 & 22.588 \\ \hline
PointSSIM Geometry Features~\cite{AlexiouPointSSIM} & - & $FR,\,GEO$& 0.815 & 0.769 & 0.162 & 0.837 & 10.30 \\ \hline
PointSSIM Luminance Features~\cite{AlexiouPointSSIM} & -& $FR,\,LUM$ & 0.718 & 0.677 & 0.194 & 0.840 & 10.10 \\ \hline
PSNR MSE Y~\cite{PSNRYUV} & -& $FR,\,LUM$& 0.580 & 0.550 & 0.229 & 0.903 & 22.588 \\ \hline
PSNR MSE YUV~\cite{PSNRYUV} & -& $FR,\,COL$& 0.638 & 0.567 & 0.215 & 0.907 & 22.588\\ \hline
Color Histogram~\cite{IreneHistogram} & -& $FR,\,COL$& 0.497 & 0.428 & 0.244 & 0.883 & 0.015\\ \hline
PCQM~\cite{MeynetPCQM} & -& $FR,\,GEO+LUM$ & 0.927 & 0.849 & 0.105 & 0.690 & 14.43 \\ \hline
Point 2 Distribution~\cite{javaheri2021pointtodistribution} & -& $FR,\,GEO+COL$ & 0.748 & 0.612 & 0.186 & 0.847 & 24.524\\ \hline
GraphSIM~\cite{QiYangGraphSIM2022} & -& $FR,\,GEO+COL$& 0.924 & 0.817 & 0.108 & \textbf{0.663} & 46.88 \\ \hline
MS GraphSIM~\cite{MSGraphSIM} & -& $FR,\,GEO+COL$ & 0.909 & 0.808 & 0.117 & 0.710 & 29.30 \\ \hline
$\mbox{PCM}_{\mbox{RR}}$~\cite{ViolaPCMRR} & -& $RR,\,GEO+COL$ & 0.567 & 0.493 & 0.232 & 0.860 & 124.28 \\ \hline
$\mbox{RR}_{\mbox{CAP}}$~\cite{ZhouRRCap}  & -& $RR,\,GEO+COL$ & 0.749 & 0.538 & 0.186 & 0.840 & 3.104 \\ \hline
FRSVR~\cite{watanabe2025full} & SVR & $FR,\,GEO+COL$ & 0.862 & 0.797 & 0.142 & 0.807 & 4.474  \\ \hline
\end{tabular} 
} 
\end{table*}

\subsection{Feature Analysis}\label{sec:RFEAnalysis}
The most important features of the metrics described in Section \ref{sec:Metrics} are analyzed using RFE~\cite{RFE}. The selection is conducted by recursively considering smaller sets of features. The estimator is trained on the initial set, finding the most important ones. Then, the least important features are removed from the current set. That procedure is recursively repeated on the pruned set until the desired number of features to select is eventually reached. 
To employ the RFE, a regression method needs to be used as an estimator, in order to rank the features. For this, the regression methods described previously are used.

\subsection{Feature combination models}
Once the most significant features were determined, every model was trained using combinations of features based on the ranking provided by the RFE, with the respective regression method.

The database was randomly partitioned at a ratio of 80\%:20\% for SVR training and testing, respectively. Data splitting was done at the level of the reference point clouds. Hence, it was assured that the reference or distorted versions of the same point cloud were either on the testing set or on the training set. 
After logistic fitting, the PCC and SROCC were computed for each of the ten random partitions.
During ten-fold cross validation, the mean squared error (MSE), mean absolute error (MAE) and coefficient of determination ($r^2$) were computed between the training scores and the test scores. Then, after the cross validation, we computed the mean for each indicator~\cite{MontesinosLópez2022Overfitting}. The results showed no case of overfitting, as the test values were never extremely higher than the training results, and the MSE, MAE and $r^2$ scores ratio between training and testing set are close to 1. Hence, it was concluded that the followed methodology did not lead to overfitting.
\begin{table*}[!t]
\begin{center}

\small


\caption{Metrics performance for the datasets referred to as subjective evaluations 1, 2 and 3. The metric combination models are defined in table~\ref{tab:crossValidation}}.\label{tab:allEvalResults}  
\resizebox{\linewidth}{!}{%
\begin{tabular}{@{}|l|l||c|c|c|c||c|c|c|c||c|c|c|c|@{}}
\hline
& & \multicolumn{4}{c||}{\makecell[c]{Subjective quality evaluation 1~\cite{EI2022}}} & \multicolumn{4}{c||}{Subjective quality evaluation 2~\cite{PrazeresACM2022}} &  \multicolumn{4}{c|}{Subjective quality evaluation 3~\cite{PrazeresICASSP2023}} \\ \hline
& Tested codecs& \multicolumn{4}{|c|}{V-PCC, G-PCC, RS-DLPCC, Draco}& \multicolumn{4}{|c|}{\makecell[l]{G-PCC, ADLPCC,PCC GEO CNNv2,\\PCGCv2,LUT\_SR}}& \multicolumn{4}{|c|}{V-PCC, G-PCC, T1, T2, T3} \\ \hline
\textbf{\textit{Metric}} & \textbf{\makecell[c]{Regression \\ Method}} & \textbf{\textit{PCC}} & \textbf{\textit{SROCC}}& \textbf{\textit{RMSE}} & \textbf{\textit{OR}} & \textbf{\textit{PCC}} & \textbf{\textit{SROCC}}& \textbf{\textit{RMSE}} & \textbf{\textit{OR}} & \textbf{\textit{PCC}} & \textbf{\textit{SROCC}}& \textbf{\textit{RMSE}} & \textbf{\textit{OR}}   \\ \hline
 \hline
PSNR MSE D1~\cite{Dtian}   & - & 0.890 & 0.884 & 0.148 & 0.618 & 0.806 & 0.782 & 0.184 & 0.753 & 0.741 & 0.725 & 0.226 & 0.781  \\ \hline
PSNR MSE D2~\cite{Dtian}   & - & 0.851 & 0.847 & 0.169 & 0.608  & 0.821 & 0.796 & 0.177 & 0.813& 0.783 & 0.773 & 0.210 & 0.769  \\ \hline
PSNR MSE Y~\cite{PSNRYUV}   & -& 0.770 & 0.772 & 0.205 & 0.688 & 0.627 & 0.617 & 0.242 & 0.767& 0.828 & 0.808 & 0.190 & 0.719  \\  \hline
PSNR MSE YUV~\cite{PSNRYUV}  & -& 0.670 & 0.679 & 0.240 & 0.719 & 0.636 & 0.658 & 0.240 & 0.820& 0.830 & 0.806 & 0.188 & 0.744 \\ \hline
Color Histogram~\cite{PSNRYUV} & -&  0.890 & 0.897 & 0.135 & 0.631 & 0.832 & 0.830 & 0.172 & 0.733& 0.680 & 0.641 & 0.249 & 0.831 \\ \hline
PCQM~\cite{MeynetPCQM}   & -& 0.944 & 0.928 & 0.106 & 0.480  & 0.899 & 0.903 & 0.137 & 0.573 & 0.873 & 0.826 & 0.167 & 0.694  \\ \hline
Point 2 Distribution~\cite{javaheri2021pointtodistribution} & -& 0.778 & 0.794 & 0.204 & 0.747 & 0.851 & 0.828 & 0.164 & 0.640& 0.866 & 0.833 & 0.169 & 0.794  \\ \hline
$\mbox{PCM}_{\mbox{RR}}$~\cite{ViolaPCMRR}  & -& 0.890 & 0.871 & 0.147 & 0.529  & 0.834 & 0.834 & 0.172 & 0.727& 0.837 & 0.831 & 0.185 & 0.763  \\ \hline
$\mbox{RR}_{\mbox{CAP}}$~\cite{ZhouRRCap}  & -& 0.718 & 0.685 & 0.226 & 0.833 & 0.735 & 0.734 & 0.212 & 0.867& 0.813 & 0.822 & 0.197 & 0.675  \\ \hline
PointSSIM~\cite{AlexiouPointSSIM}   & -& 0.869 & 0.867 & 0.160 & 0.588 & 0.859 & 0.857 & 0.159 & 0.720 & 0.706 & 0.684 & 0.239 & 0.831   \\ \hline
GraphSIM~\cite{QiYangGraphSIM2022}  & -& 0.907 & 0.893 & 0.137 & 0.500 & 0.800 & 0.799 & 0.186 & 0.780& 0.919 & 0.900 & 0.135 & 0.569  \\ \hline
MS-GraphSIM~\cite{MSGraphSIM}   & -& 0.902 & 0.880 & 0.179 & 0.490 & 0.890 & 0.884 & 0.142 & 0.620 & \textbf{0.925} & \textbf{0.901} & \textbf{0.130} & \textbf{0.500}  \\ \hline
FRSVR~\cite{watanabe2025full} & SVR &  0.811 & 0.763 & 0.189 & 0.686 & 0.780 & 0.655 & 0.194 & 0.753 & 0.679 & 0.661 & 0.247 & 0.838 \\ \hline
\hline
Model 1 (8 features) & SVR & 0.943 & 0.917 & 0.107 & 0.589 & 0.903 & 0.857 & 0.133 & 0.693 & 0.906 & 0.860 & 0.143 & 0.675 \\ \hline
Model 2 (10 features) & SVR & 0.945 &	0.913 &	0.105 &	0.520 &	0.889 &	0.834 &	0.143 &	0.720 &	0.907 &	0.855 &	0.143 &	0.688 \\ \hline
Model 3 (14 features)  & SVR & 0.938 &  0.871 &  0.112 &  0.529 &0.884 &  0.814 &  0.145 &  0.753 &0.905 &  0.853 &  0.144 &  0.681\\ \hline
Model 4 (4 features) &  SVR & 0.961 & 0.937 & 0.089 & 0.490  & 0.924 & 0.888 & 0.118 & 0.707 & 0.906 & 0.850 & 0.143	& 0.675 \\ \hline
Model 5 (6 features) (FSM)  & RR & \textbf{0.958} & 0\textbf{.939 }& \textbf{0.093} & 0.490 & 0.916 & 0.908 & 0.123 & 0.653 & 0.909 & 0.878 & 0.142 & 0.625  \\ \hline
Model 6 (11 features) & RR & 0.951 & 0.937 & 0.100 & 0.510 & 0.919 & 0.913 & 0.120 & 0.640 & 0.897 & 0.872 & 0.150 & 0.600 \\ \hline
Model 7 (15 features) & RR & 0.947 & 0.935 & 0.104 & 0.480 & 0.913 & 0.907 & 0.124 & 0.633 & 0.904 & 0.880 & 0.145 & 0.613 \\ \hline
Model 8 (4 features)  & RR & 0.956 & 0.938 & 0.095 & \textbf{0.422} & \textbf{0.930} & \textbf{0.926} & \textbf{0.113} & \textbf{0.553} & 0.894 & 0.866 & 0.153 & 0.606  \\ \hline

\end{tabular}%
}
\end{center}
\end{table*}

\begin{table*}
\caption{Metrics performance for Waterloo and SJTU-PCQA. The metric combination models are defined in table~\ref{tab:crossValidation}}.\label{tab:allDatabaseResults}  
\begin{center}
\small
\begin{tabular}{@{}|l|c||c|c|c|c||c|c|c|c|@{}}
\hline
 & & \multicolumn{4}{c||}{Waterloo~\cite{LIUQI2022}} & \multicolumn{4}{c|}{SJTU-PCQA~\cite{projectionYANG}} \\ \hline
\textbf{\textit{Metric}} & \textbf{\makecell[c]{Regression \\ Method}} & \textbf{\textit{PCC}} & \textbf{\textit{SROCC}}& \textbf{\textit{RMSE}} & \textbf{\textit{OR}} & \textbf{\textit{PCC}} & \textbf{\textit{SROCC}}& \textbf{\textit{RMSE}} & \textbf{\textit{OR}}   \\ \hline
 \hline
PSNR MSE D1~\cite{Dtian} & - & 0.578 & 0.566 & 0.203 & 0.935& 0.873 & 0.798 & 0.135 & \textit{0.802}  \\ \hline
PSNR MSE D2~\cite{Dtian} & - & 0.481 & 0.461 & 0.219 & 0.932 & 0.762 & 0.678 & 0.180 & 0.830  \\ \hline
PSNR MSE Y~\cite{PSNRYUV} & - & 0.608 & 0.587 & 0.197 & 0.939& 0.701 & 0.704 & 0.197 & 0.892  \\ \hline
PSNR MSE YUV~\cite{PSNRYUV} & - &  0.551 & 0.536 & 0.207 & 0.935 &  0.655 & 0.659 & 0.211 & 0.923  \\ \hline
Color Histogram~\cite{PSNRYUV} & - & 0.195 & 0.199 & 0.243 & 0.951& 0.068 & 0.111 & 0.280 & 0.934  \\ \hline
PCQM~\cite{MeynetPCQM} & - &  0.750 & 0.743 & 0.165 & 0.884 &  0.858 & 0.844 & 0.142 & 0.812  \\ \hline
Point 2 Distribution~\cite{javaheri2021pointtodistribution}  & - & 0.462 & 0.432 & 0.222 & 0.932& 0.632 & 0.620 & 0.217 & 0.881  \\ \hline
$\mbox{PCM}_{\mbox{RR}}$~\cite{ViolaPCMRR} & - & 0.368 & 0.345 & 0.232 & 0.931 & - & - & - & -  \\ \hline
$\mbox{RR}_{\mbox{CAP}}$~\cite{ZhouRRCap}   & - & 0.708 & 0.715 & 0.176 & 0.936 & 0.765 & 0.752 & 0.180 & 0.899  \\ \hline
PointSSIM~\cite{AlexiouPointSSIM} & - & 0.468 & 0.455 & 0.220 & 0.928 & 0.723 & 0.705 & 0.191 & 0.910 \\ \hline
GraphSIM~\cite{QiYangGraphSIM2022} & - & 0.690 & 0.681 & 0.180 & 0.918 &  0.868 & 0.854 & 0.138 & 0.820  \\ \hline
MS-GraphSIM~\cite{MSGraphSIM} & - & 0.716 & 0.708 & 0.174 & 0.914 & \textbf{0.893} & \textbf{0.874} & \textbf{0.125} & 0.831  \\ \hline
FRSVR~\cite{watanabe2025full} & SVR & 0.391 & 0.181 & 0.228 & 0.949 & 0.606 & 0.614 & 0.220 & 0.902 \\ \hline
\hline
Model 1 (8 features) & SVR & 0.676 & 0.686 & 0.183 & 0.892 & 0.872 & 0.856 & 0.136 & 0.836  \\ \hline
Model 2 (10 features) & SVR & 0.676 & 0.680 & 0.183 & 0.908 & 0.879 & 0.859 & 0.132 & 0.839   \\ \hline
Model 3 (14 features)  & SVR & 0.679 &  0.688 &  0.183 &  0.916 &0.889 &  0.865 &  0.127 &  \textbf{0.786} \\ \hline
Model 4 (4 features) & SVR & 0.758 & 0.760 & 0.162 & 0.896 & 0.882 & 0.840 & 0.130 & 0.815   \\ \hline
Model 5 (6 features) (FSM) & RR& 0.702& 0.715& 0.177& 0.896& 0.889& 0.870& 0.127&0.825\\ \hline
Model 6 (11 features)& RR& 0.686& 0.681& 0.247& 0.947 & 0.856& 0.842& 0.143& 0.841\\ \hline
Model 7 (15 features)& RR& 0.686& 0.690& 0.181& 0.912& 0.874& 0.857& 0.134& 0.815\\ \hline
Model 8 (4 features)& RR& \textbf{0.788}& \textbf{0.790}& \textbf{0.153}& \textbf{0.884}& 0.881& 0.868& 0.131& 0.823\\ \hline
\end{tabular}%
\end{center} 
\end{table*}
Furthermore, the standard deviation ($\sigma$) of both PCC and SROCC was also computed to understand if there were large variations between each split. 
The results are shown in table \ref{tab:crossValidation}.
The table shows the combination of metrics and features that lead to the best results.
Between them, there were slight performance changes, and the models achieved a similar correlation for both PCC and SROCC.
It is noted that the best performance is achieved by model 7 considering features from PCQM, MS-GraphSIM, PSNR MSE D2, PSNR MSE YUV, and PointSSIM ($PCC/SROCC=0.951/0.9$).
It can also be observed that the standard deviations of the PCC and SROCC  are quite small, indicating that each fold had a similar performance.
Nonetheless, it is noted that the best performance is achieved by model 7 considering features from PCQM, MS-GraphSIM, PSNR MSE D2, PSNR MSE YUV, and PointSSIM ($PCC/SROCC=0.951/0.9$), using RR as a regression method. It is closely followed by model 3. That model considers features defined in the same metrics, but trained with an SVR.

Table \ref{tab:Validation} shows the quality features combination models performance on the BASICS validation dataset, after training the models with the complete training dataset. The validation dataset contains 300 distorted point clouds, coded with the same codecs from the training dataset, from 15 reference point clouds.
Moreover, the results of the state-of-the-art metrics in this database are also shown for comparison purposes. 
The second column of tables \ref{tab:crossValidation} and \ref{tab:Validation} defines if the metric is full-reference ($FR$), reduced-reference ($RR$), or no-reference ($NR$). Furthermore, it is also described if the metric considers only the geometry ($GEO$), color ($COL$), luminance ($LUM$), both geometry and color ($GEO+COL$) information or both geometry and luminance ($GEO+LUM$) information. $COL$  means that the selected features use both chromatic and luminance information. The final column shows the average time each metric took to compute, for the BASICS dataset. As expected, the computational complexity grows if  more features are added.

It can be observed that the best PCC value is obtained using model 5, that contains features defined in PCQM and MS-GraphSIM, and the PSNR MSE D2, using the RR regression algorithm, and the best SROCC is achieved by model 1, containing features defined in PCQM and MS-GraphSIM, and the PSNR MSE D2, using an SVR. The obtained RMSE and OR are also quite low, when compared to the other state-of-the-art-metrics, although the best OR value is obtained by GraphSIM. It is also observed that the obtained results are different from the obtained in the training dataset.
\begin{figure}
    \centering
    \subfigure[SVR]{\label{SVRHist}\includegraphics[width=0.5\linewidth]{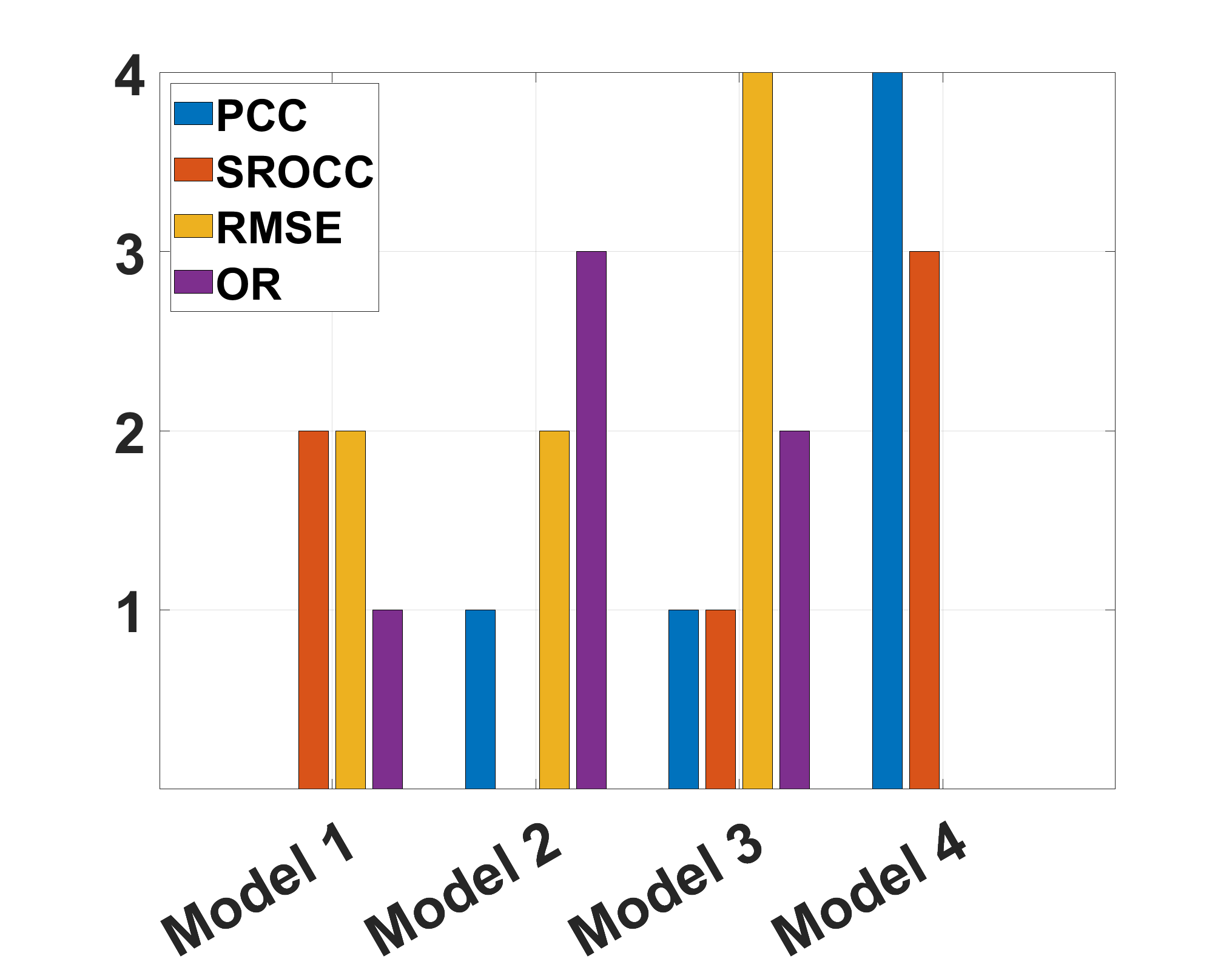}}\hfill 
    \subfigure[RR]{\label{RRHist}\includegraphics[width=0.5\linewidth]{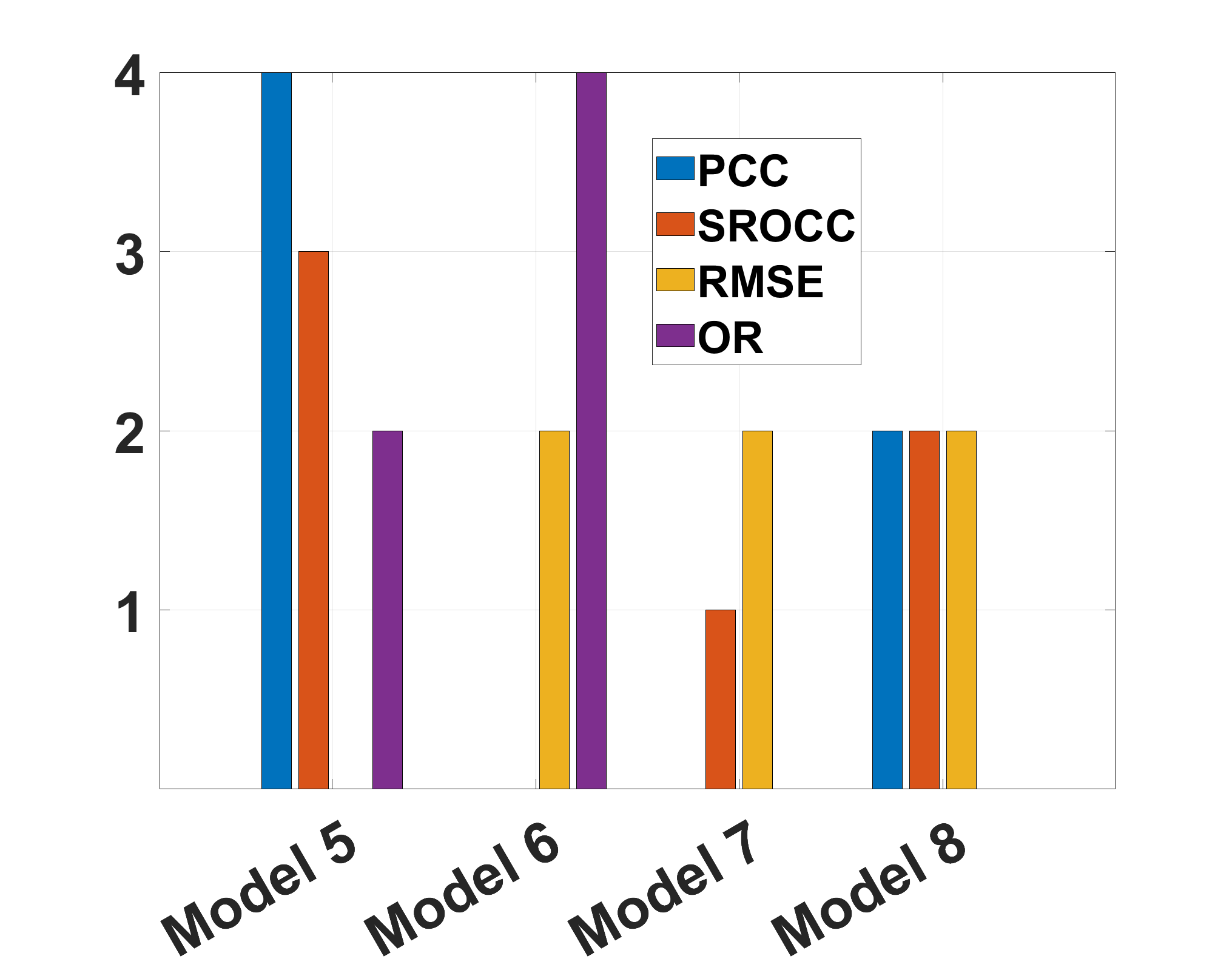}}\hfill
     \caption{Histograms representing the number of times that each model performs the best for PCC, SROCC, RMSE and OR.}
    \label{fig:histogram}
\end{figure}

\begin{figure*}[!b]
    \centering
    \subfigure[GraphSIM]{\label{GraphEI}\includegraphics[width=0.3\linewidth]{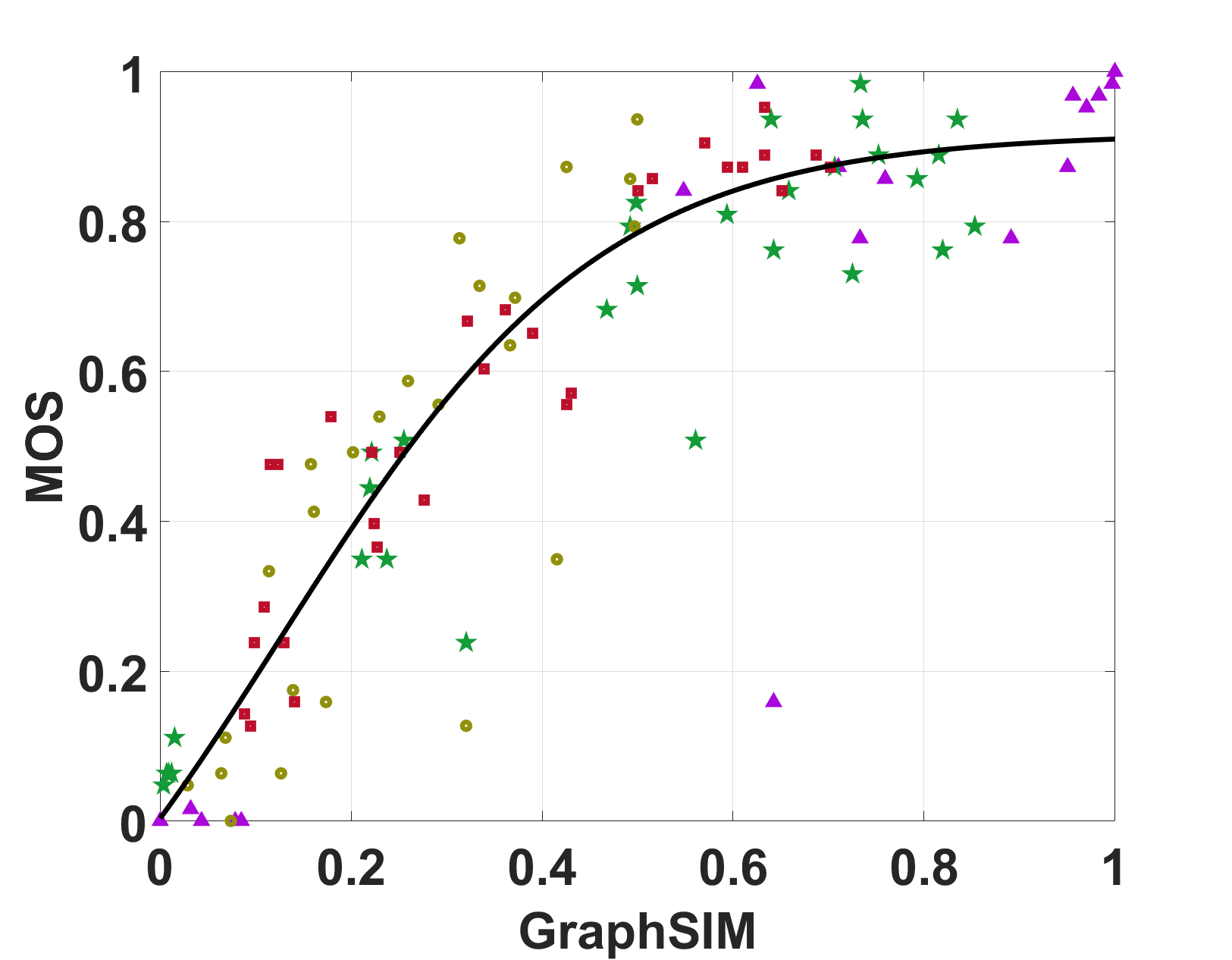}}\qquad
    \subfigure[MS-GraphSIM]{\label{MSGraphEI}\includegraphics[width=0.3\linewidth]{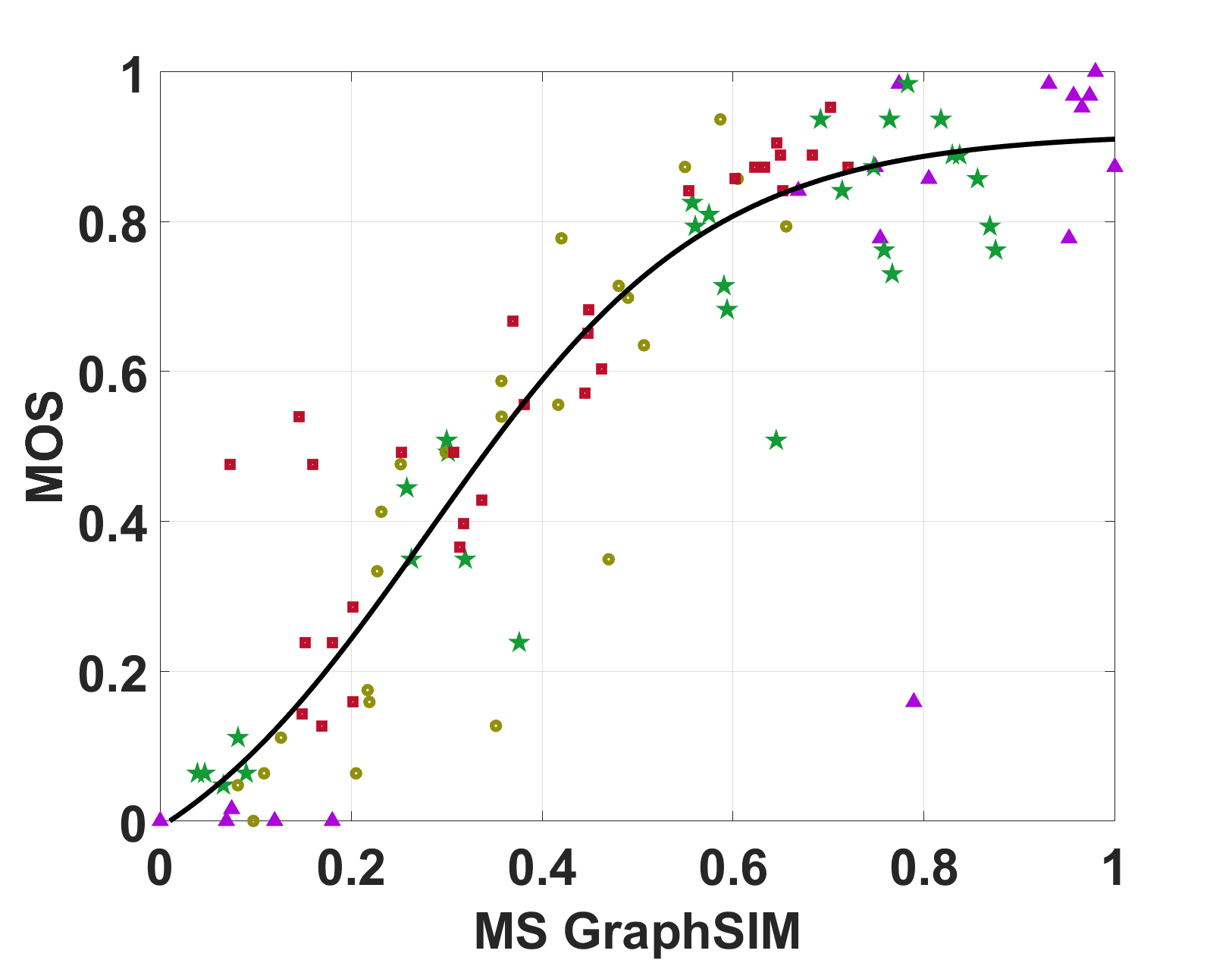}}\qquad 
    \subfigure[PCQM]{\label{PCQMEI}\includegraphics[width=0.3\linewidth]{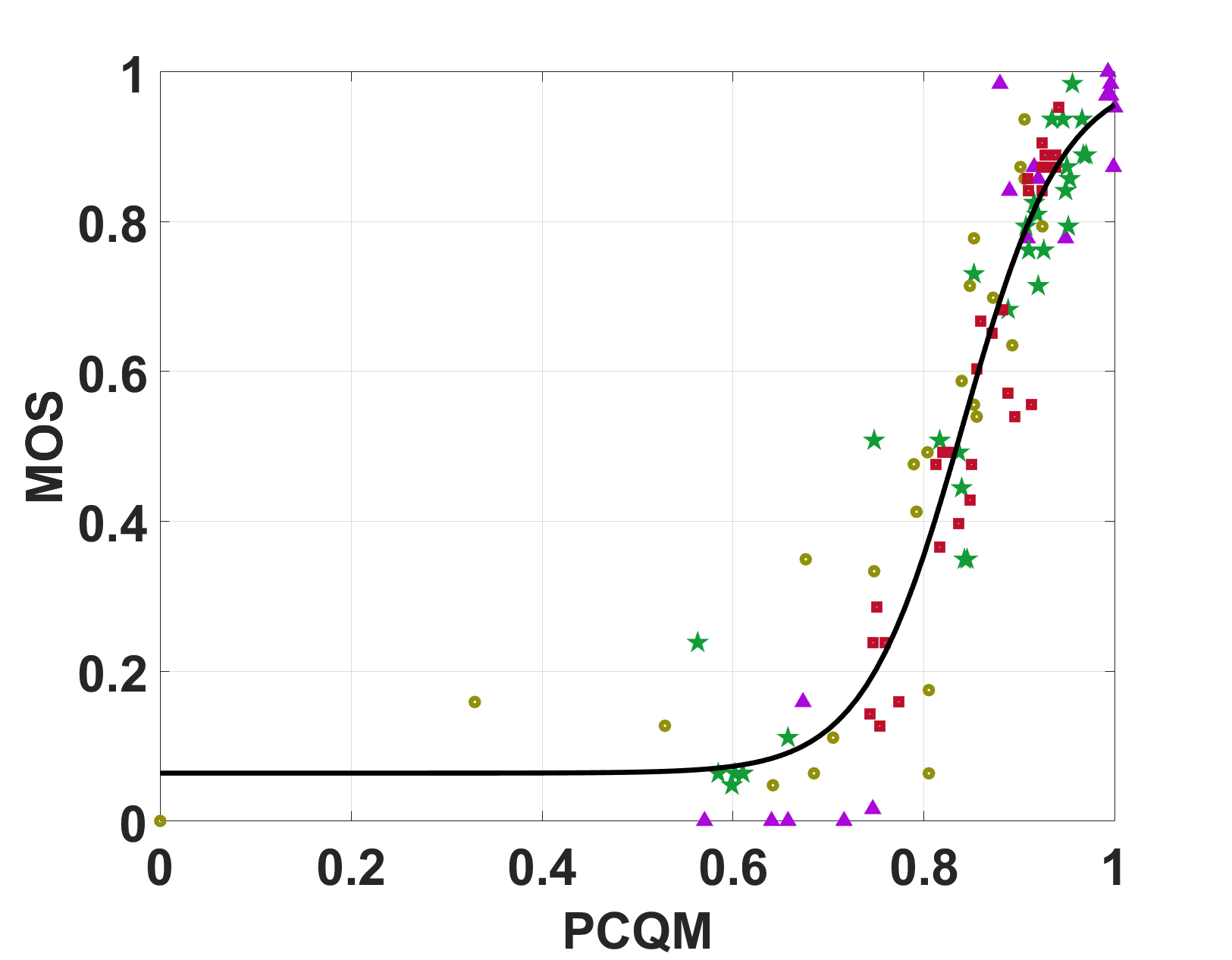}}\\
    \subfigure[Model 4]{\label{Model4MEI}\includegraphics[width=0.3\linewidth]{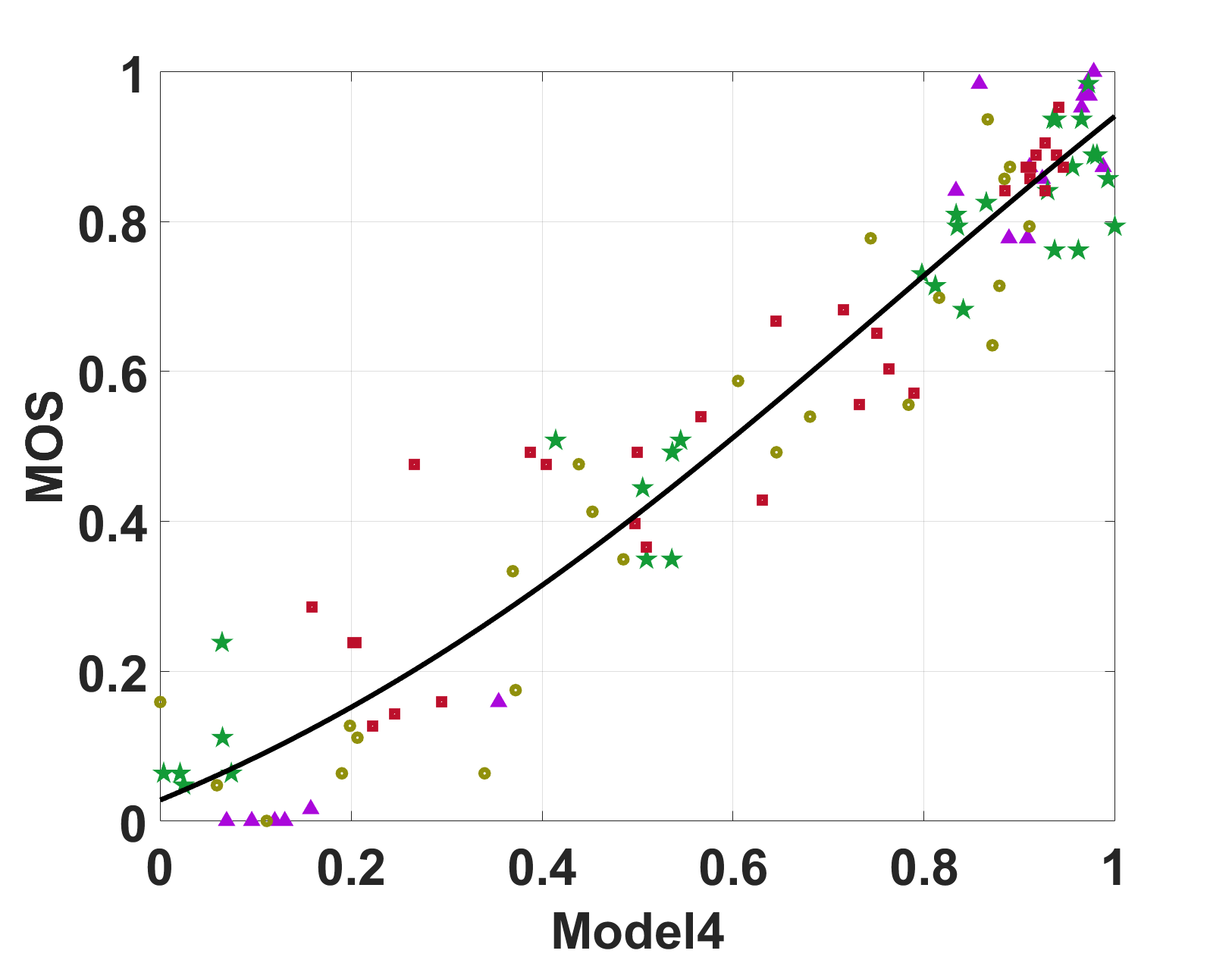}}\qquad
    \subfigure[Model 5 (FSM)]{\label{Model5EI}\includegraphics[width=0.3\linewidth]{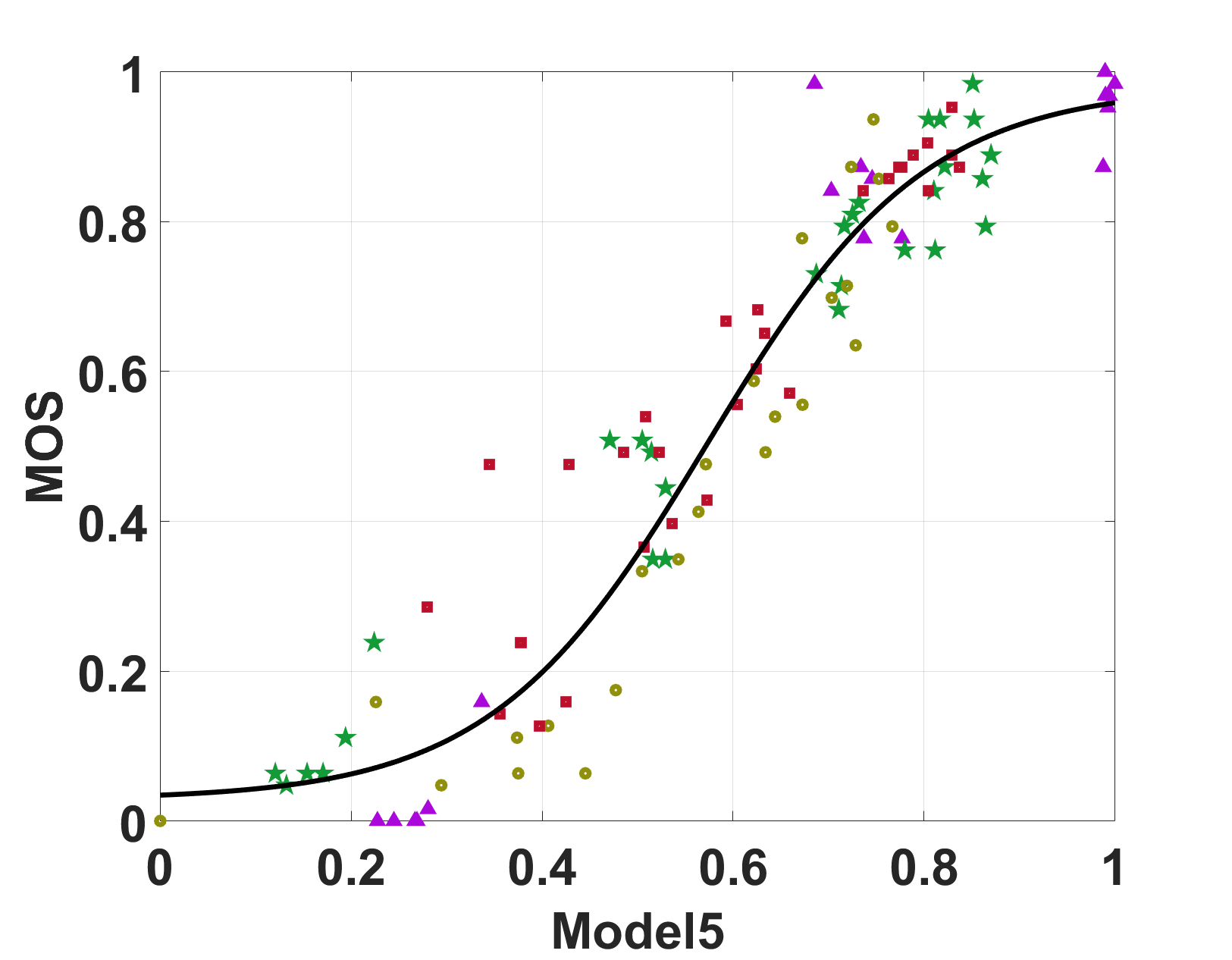}}\\
    \centering
    \includegraphics[width=0.32\linewidth]{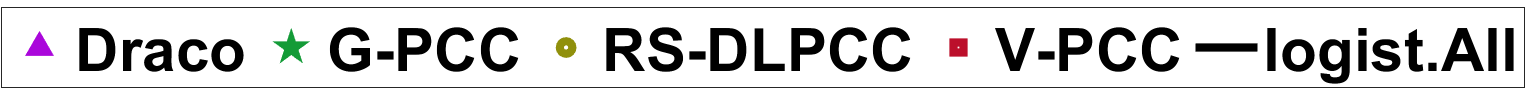}    
    \caption{Metric vs. MOS plots, with logistic regression curves of FSM and the three best performing metrics for subjective quality evaluation 1. \cite{EI2022}.}
    \label{LogisticFitting}
\end{figure*}

\section{Evaluating Feature Combinations} \label{sec:Validation}

Objective quality metrics should be validated using subjective quality evaluation results as ground truth.
As such, the default implementation of the state-of-the-art full-reference metrics PSNR MSE D1, PSNR MSE D2, PSNR MSE Y, PSNR MSE YUV, Point 2 Distribution, PointSSIM, PCQM, GraphSIM, Color Histogram, MS-GraphSIM, FRSVR and the reduced-reference metrics $\mbox{PCM}_{\mbox{RR}}$ and $\mbox{RR}_{\mbox{CAP}}$ were selected. A number of subjective quality evaluations were conducted in order to validate the models.
It was ensured that none of the point clouds present in those evaluations were present in the training set. Subjective quality evaluation 1 reported on a subjective evaluation~\cite{EI2022} comparing V-PCC and G-PCC using the octree mode, the deep-learning solution RS-DLPCC~\cite{GuardaRSDLPCC} and the Draco(https://github.com/google/draco) codec. 
Subjective quality evaluation 2, which is focused on learning-based coding solutions~\cite{PrazeresACM2022} quality evaluation, was also chosen. In particular, the learning-based codecs ADLPCC~\cite{GuardaDL}, GeoCNNv2~\cite{quach2020improved}, and PCGCv2~\cite{Jianqiang-PCGCv2} performance was analyzed. 
The LUT\_SR~\cite{QueirozLut} solution was also considered. The codecs were compared to the octree mode of G-PCC.
Finally, subjective quality evaluation 3, reports on the results of the JPEG Pleno Call for Proposals~\cite{PrazeresICASSP2023}. The call focused on deep-learning solutions. As the call is anonymous, the solutions are referred to as T1, T2, and T3, and they all can encode both geometry and color information of point cloud static content. 
The codecs were compared against G-PCC and V-PCC.

All chosen subjective quality evaluations consider deep-learning based solutions. This is important as these types of codecs are becoming increasingly popular, and extensive research is being conducted on them. 
For that reason, it is important to verify how the objective quality evaluation models perform when evaluating the performance of these codecs.
Furthermore, two popular point cloud quality assessment (PCQA) datasets were considered, notably the Waterloo~\cite{LIUQI2022} and the Shanghai Jiao Tong University point cloud quality assessment (SJTU-PCQA)~\cite{projectionYANG}.
The Waterloo database contains 700 point clouds coded with V-PCC, G-PCC (octree and trisoup modes), Downsampling and Gaussian noise, targeting both texture and geometry distortions. SJTU-PCQA contains 378 point clouds coded with octree-based compression, color noise, downscaling, downscaling and color noise, downscaling and geometry gaussian noise, geometry gaussian noise, color noise, and geometry gaussian noise.
For the considered metrics and feature combination models, the statistical measures proposed in~\cite{ITU-T} were computed, specifically the PCC, the SROCC, the Root Mean Squared Error (RMSE) and the Outlier Ratio (OR). The MOS predicted for each of the objective metrics was calculated by applying a logistic fit function to the objective scores, as is commonly done when benchmarking objective metrics~\cite{HDRMarco,LazzarottoMMSP2021}. All the combinations in table~\ref{tab:crossValidation} were evaluated, as they show similar performance.
The results are shown in tables \ref{tab:allEvalResults} and \ref{tab:allDatabaseResults}. 

Analyzing the results for the subjective quality evaluation 1~\cite{EI2022}, it can be observed that the defined models that combine several features achieve better performance than the individual metrics. It should be noted that the models have a very similar performance for both PCC and SROCC as well as RMSE and OR, independently of the regression method. The best performing model for this evaluation is model 5 (6 features with RR).

Regarding subjective quality evaluation 2~\cite{PrazeresACM2022}, it can be observed that models 8 (4 features with RR) achieves the best correlation values. 
Furthermore, both RMSE and OR are also lower than the ones obtained for the state-of-the-art metrics for each combination model.

The subjective quality evaluation 3~\cite{PrazeresICASSP2023} provided very competitive results, although the performance provided by GraphSIM and MS-GraphSIM is slightly better. 
In this evaluation, the performance achieved by the different models is once again very similar, although the model 5 ( 6 features trained with RR)  achieves the best performance.

The results for the Waterloo and SJTU-PCQA dataset, shown in table~\ref{tab:allDatabaseResults}, reveal a lower performance. Moreover, other metrics also have a reduced performance for both databases. These databases contain a wide range of distortions including gaussian noise and downscalling, not present in the BASICS database (used for training) which causes this performance reduction.
However, even in this situation, the combination models obtained for the different regression methods generally achieve the best performance.
For the Waterloo database, model 8 (4 features with RR) shows the best performance. For the SJTU-PCQA dataset, the best performing metric is the MS-GraphSIM, closely followed by model 5 (6 Features with RR).

From the results shown in tables \ref{tab:crossValidation}, \ref{tab:allEvalResults} and \ref{tab:allDatabaseResults}, it is very difficult to select the best performing model. Most of the combination models achieved very high performance for PCC, SROCC, RMSE and OR. To assess the most consistent ones, the times that a model achieved the best PCC, SROCC, RMSE or OR values were identified, allowing to access the best model for each regression method.
Then the same comparison was conducted between the identified best models. Figure \ref{fig:histogram} shows the results for each feature combination model, and shows that models 4 and 5 are the best models for the SVR and RR regression methods, respectively.

Figure \ref{LogisticFitting} shows the plots of normalized MOS vs. metrics for each of the best models and the best metrics for the subjective quality evaluation 1.  It can be observed that the results for each model are close to the logistic curve, but the curve of model 5 reveals to be the best fit. 



\section{Conclusions}\label{sec:conclusions}
This paper reports a study on the performance of objective quality assessment features for point cloud compression models. The studied features were defined in different widely used full-reference point-cloud metrics. An analysis using RFE showed that the features defined in the metrics PCQM and MS-GraphSIM, as well as the metric PSNR MSE D2, are the most representative of subjective quality.
Based on previous studies~\cite{PRAZERESSPIC2024,ak2023basics}, it was expected that PCQM and MS-GraphSIM features would provide the most relevant features for the best performing models. Furthermore, PSNR MSE D2 reveals to be a very relevant feature for some selected models.

This study started with an analysis of the importance of each objective quality feature on the estimation of subjective quality.
Then, several combinations were evaluated, and a new metric referred to as FSM was defined, combining features from MS-GraphSIM and PCQM, and the PSNR MSE D2, using RR.
The luminance features defined in GraphSIM and the luminance features defined in PCQM revealed a larger impact on the prediction of the quality of compressed point clouds. The PSNR MSE D2 was also revealed to be one of the most representative feature for quality estimation using the BASICS dataset.
However, this result might be influenced by the importance of that measure in the optimization process of the codecs used for the generation of the distorted point clouds in this dataset.
In particular, PSNR MSE D2 is used in the optimization of the deep-learning-based codec GeoCNN.
The FSM showed increased performance over both PCQM and MS-GraphSIM in most evaluations and datasets.
Across five datasets, FSM achieved results similar to the most popular full reference metrics, showing consistent results across all evaluations with correlations greater than 0.9 for most of the datasets. 
Furthermore, the model also provides a good representation of the quality scores when deep learning codecs are present, which was an identified issue for the typical full-reference metrics found in previous studies that motivated the current study.

\bibliographystyle{IEEEtran}
\bibliography{references}
\vspace{-1.5cm}
\begin{IEEEbiography}
[{\includegraphics[width=1in,height=1.25in,clip,keepaspectratio]{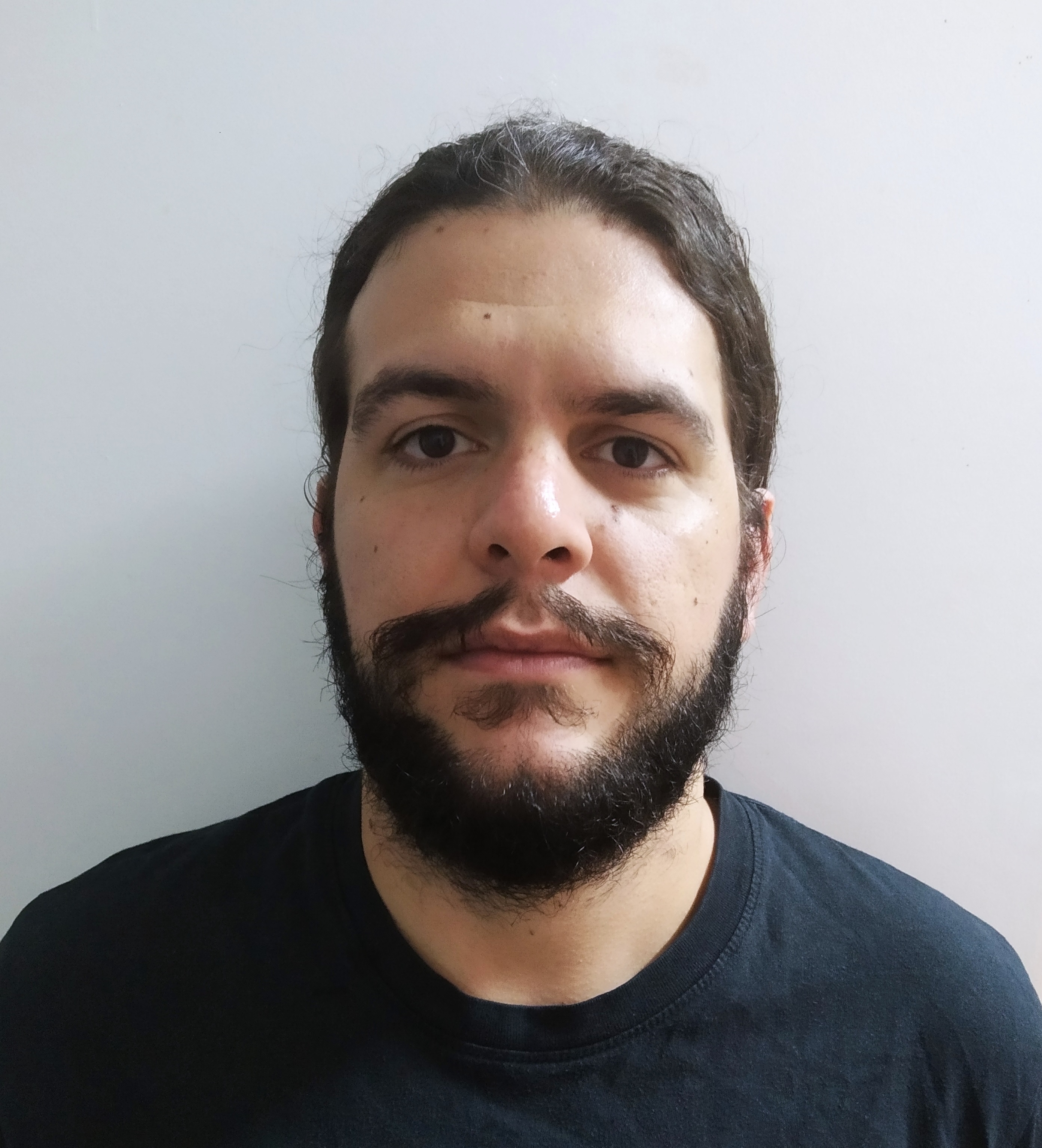}}]{João Prazeres}  (IEEE Student Member) is a PhD candidate from Universidade da Beira Interior (UBI), Covilhã. He graduated in Electrical and computer engineering in Universidade da Beira Interior in 2018 and received his master degree in 2020. 
He has been deeply involved in the JPEG PLENO Point Cloud Coding activity.
Recently he received a best paper award in 3D Imaging and Applications of the Electronic Imaging Symposium 2022. \\
\end{IEEEbiography}
\vspace{-1.5cm}
\begin{IEEEbiography}
[{\includegraphics[width=1in,height=1.25in,clip,keepaspectratio]{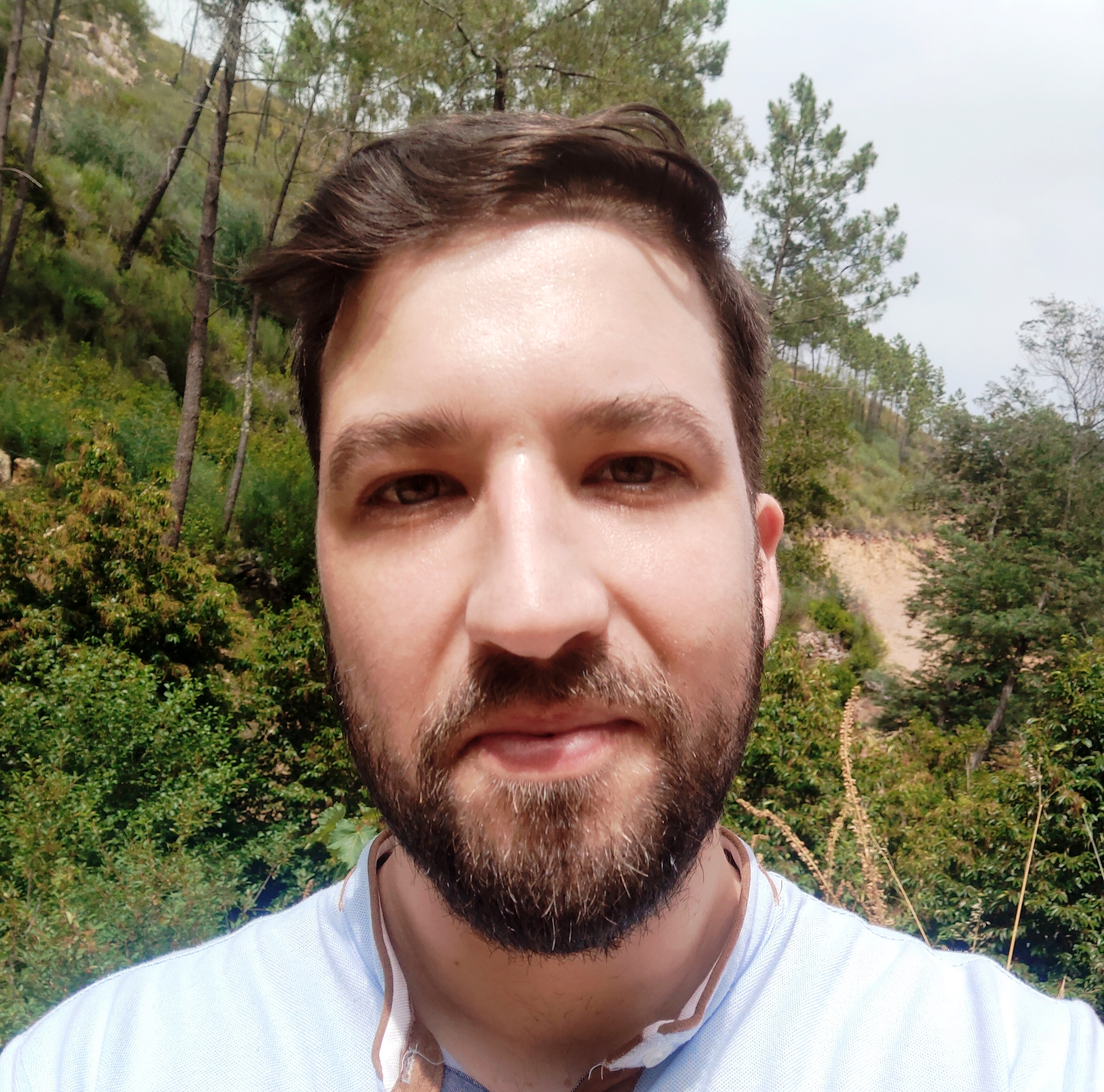}}]{Rafael Rodrigues} (IEEE Student Member) is a Ph.D. candidate in Electrical and Computer Engineering at Universidade da Beira Interior (UBI), Portugal. He received his B.Sc. in Biomedical Sciences and M.Sc. in Electrical and Computer Engineering, both from UBI. His primary research interests are on medical image processing and computer-aided diagnosis, with involvement on the EU COST Action BM1304 - MYO-MRI. However, he has also collaborated closely in many research efforts on image and video coding and quality assessment. He is currently a Portuguese JPEG member, as well as a Qualinet (COST IC1003) and VQEG member.
\end{IEEEbiography}
\vspace{-1.5cm}
\begin{IEEEbiography}
[{\includegraphics[width=1in,height=1.25in,clip,keepaspectratio]{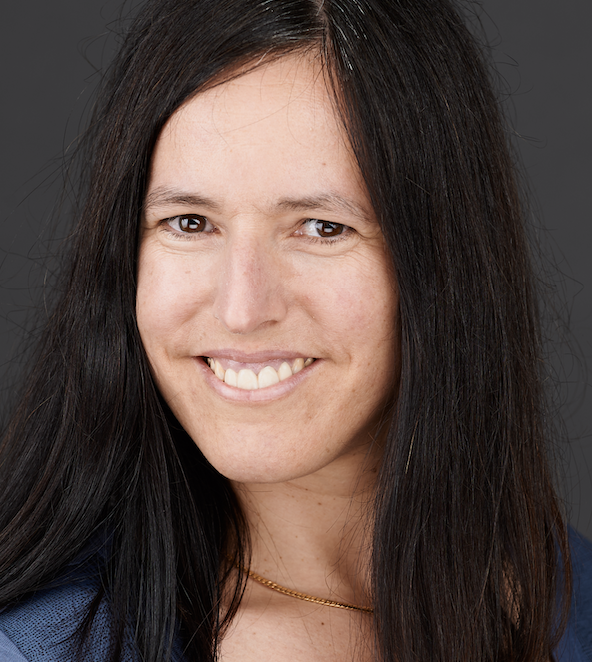}}]{Manuela Pereira}
received the 5-year B. S. degree in Mathematics and Computer Science in 1994 and the M. Sc. degree in Computational Mathematics in 1999, both from the University of Minho, Portugal. She received the Ph. D. degree in Signal and Image Processing in 2004 from the University of Nice Sophia Antipolis, France. Currently she is the Head of the Multimedia Signal Processing Group from Instituto de Telecomunicacoes at the University of Beira Interior. She is an Associate Professor in the Computer Science Department 
of the University of Beira Interior, Portugal. Her main research interests include: Image and Video Coding; Multimedia technologies standardization; 
Signal Processing for Telecommunications; Information theory; Real-time video streaming; 3D and 4D Imaging; Medical Imaging.
\end{IEEEbiography}
\vspace{-1.5cm}
\begin{IEEEbiography}
[{\includegraphics[width=1in,height=1.25in,clip,keepaspectratio]{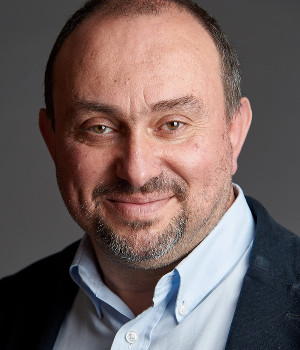}}]{António M.G Pinheiro}
(M'99, SM'15)
Is an Associate Professor at UBI (Universidade da Beira Interior), and a researcher at IT (Instituto de Telecomunicações), Portugal. He received the "Licenciatura" in Electrical and Computer Engineering from IST, Lisbon in 1988 and the PhD in Electronic Systems Engineering from University of Essex, UK in 2002. 
He is a Portuguese delegate to ISO/IEC JTC1/SC29 and the Communication Subgroup chair of JPEG. 
He was the PC co-chair of QoMEX 2015, special session co-chair of QoMEX 2016, 
and organiser of the tutorial in ACM Multimedia 2021 ”Plenoptic Quality Assessment: The JPEG Pleno Experience”.
He is Associate editor of IEEE Trans. on Multimedia and a senior member of IEEE. \end{IEEEbiography}

\end{document}